\documentclass[10pt]{article}

\usepackage{amsmath}
\usepackage{amssymb}

\usepackage{graphicx}

\usepackage{cite}
\usepackage{cancel}

\usepackage{color} 

\usepackage[labelfont=bf,labelsep=period,justification=raggedright]{caption}

\makeatletter
\renewcommand{\@biblabel}[1]{\quad#1.}
\makeatother

\date{}

\usepackage{amsthm}
\usepackage{thmtools}
\usepackage{thm-restate}
\usepackage{nameref}

\newcommand{\rcs}[1]{\textcolor{blue}{[rephrase/compress phrasing]}}

\newcommand{\rp}[1]{\textcolor{blue}{[rephrase wording]}}

\newcommand{\xxi}{\boldsymbol{\xi}}
\newcommand{\eeta}{\boldsymbol{\eta}}
\newcommand{\zzeta}{\boldsymbol{\zeta}}
\newcommand{\cchi}{\boldsymbol{\chi}}
\newcommand{\mmu}{\boldsymbol{\mu}}
\newcommand{\iin}{I_{x}}
\newcommand{\iout}{I_{y}}
\newcommand{\itot}{I^{\text{tot}}}
\newcommand{\ieta}{I_\eta}
\newcommand{\id}{{\bf I}}
\newcommand{\ff}{{\bf f}}
\newcommand{\rr}{{\bf r}}
\newcommand{\hh}{{\bf h}}
\newcommand{\uu}{{\bf u}}
\newcommand{\GG}{\overline{{\bf G}}}
\newcommand{\fprime}{{\bf f}'}
\newcommand{\oo}{{\boldsymbol \Omega}}
\newcommand{\pp}{{\bf P}}
\newcommand{\ww}{{\bf W}}
\newcommand{\Sigeff}{{\boldsymbol \Sigma}_{\text{eff},\eta}}
\newcommand{\Sigtot}{{\boldsymbol \Sigma}_\text{tot}}
\newcommand{\Sig}{\boldsymbol{\Sigma}}
\newcommand{\Sige}{{\boldsymbol \Sigma}_\eta}
\newcommand{\Sigy}{{\boldsymbol \Sigma}_y}
\newcommand{\Sigyeff}{{\boldsymbol \Sigma}_{\text{eff},y}}
\newcommand{\Sigx}{{\boldsymbol \Sigma}_\xi}
\newcommand{\weff}{{\bf W}_{\text{eff}}}

\newcommand{\xx}{{\bf x}}
\newcommand{\yy}{{\bf y}}

\newcommand{\order}{{\mathcal O}}

\newcommand{\Var}{\text{Var}}
\newcommand{\Cov}{\text{Cov}}

\newcommand{\eps}{\epsilon}

\newcommand{\beq}{\begin{equation}}
\newcommand{\eeq}{\end{equation}}
\newcommand{\barr}{\begin{array}}
\newcommand{\earr}{\end{array}}
\newcommand{\beqr}{\begin{eqnarray}}
\newcommand{\eeqr}{\end{eqnarray}}
\newcommand{\beqrn}{\begin{eqnarray*}}
\newcommand{\eeqrn}{\end{eqnarray*}}
\newcommand{\beqn}{\begin{equation*}}
\newcommand{\eeqn}{\end{equation*}}
\newcommand{\bei}{\begin{itemize}}
\newcommand{\beii}{\begin{itemize} \item}
\newcommand{\eei}{\end{itemize}}
\newcommand{\ben}{\begin{enumerate}}
\newcommand{\een}{\end{enumerate}}
\newcommand{\bes}{\begin{small}}
\newcommand{\ees}{\end{small}}
\newcommand{\bec}{\begin{center}}
\newcommand{\eec}{\end{center}}
\newcommand{\betab}{\begin{tabular}}
\newcommand{\eetab}{\end{tabular}}

\theoremstyle{definition}

\theoremstyle{remark}

\newcommand{\tr}{\mathrm{tr}}

\makeatletter
\renewcommand{\@biblabel}[1]{\quad#1.}
\makeatother

\begin{document}
\begin{flushleft}
{\Large
\textbf{Robust information propagation through noisy neural circuits}
}
\\
\smallskip
Joel Zylberberg$^{1,2,3,4,\ast}$,
Alexandre Pouget$^{5,6}$,
Peter E. Latham$^{6,\dagger}$,
Eric Shea-Brown$^{3,7,8,\dagger}$
\smallskip
\\
\textbf{1} Department of Physiology and Biophysics, Center for Neuroscience, and Computational Bioscience Program, University of Colorado School of Medicine, Aurora, Colorado, United States of America
\\
\textbf{2} Department of Applied Mathematics, University of Colorado, Boulder, Colorado, United States of America
\\
\textbf{3} Department of Applied Mathematics, University of Washington, Seattle, Washington, United States of America
\\
\textbf{4} Learning in Machines and Brains Program, Canadian Institute For Advanced Research, Toronto, Ontario, Canada
\\
\textbf{5} Department of Basic Neuroscience, University of Geneva, Switzerland
\\
\textbf{6} Gatsby Computational Neuroscience Unit, University College London, London, United Kingdom
\\
\textbf{7} Department of Physiology and Biophysics, Program in Neuroscience, University of Washington Institute for Neuroengineering, and Center for Sensorimotor Neural Engineering, University of Washington, Seattle, Washington, United States of America\\
\textbf{8} Allen Institute for Brain Science, Seattle, Washington, United States of America
\\
$\dagger$ These authors contributed equally to this work
\\
$\ast$ Email: joel.zylberberg@ucdenver.edu
\end{flushleft}

\section*{Abstract} 
Sensory neurons give highly variable responses to stimulation, which can limit the amount of stimulus information available to downstream circuits. Much work has investigated the factors that affect the amount of information encoded in these population responses, leading to insights about the role of covariability among neurons, tuning curve shape, etc. However, the informativeness of neural responses is not the only relevant feature of population codes; of potentially equal importance is how robustly that information propagates to downstream structures. For instance, to quantify the retina's performance, one must consider not only the informativeness of the optic nerve responses,
but also the amount of information that survives the spike-generating nonlinearity and noise corruption in the next stage of processing, the lateral geniculate nucleus. Our study identifies the set of covariance structures for the upstream cells that optimize the ability of information to propagate through noisy, nonlinear circuits. Within this optimal family are covariances with ``differential correlations'', which are known to reduce the information encoded in neural population activities. Thus, covariance structures that maximize information in neural population codes, and those that maximize the ability of this information to propagate, can be very different. Moreover, redundancy is neither necessary nor sufficient to make population codes robust against corruption by noise: redundant codes can be very fragile, and synergistic codes can -- in some cases -- optimize robustness against noise.

\section*{Author Summary}
Information about the outside world, which originates in sensory neurons, propagates through multiple stages of processing before reaching the neural structures that control behavior. While much work in neuroscience has investigated the factors that affect the amount of information contained in peripheral sensory areas, very little work has asked how much of that information makes it through subsequent processing stages. That's the focus of this paper, and it's an important issue because information that fails to propagate cannot be used to affect decision-making. We find a tradeoff between information content and information transmission: neural codes which contain a large amount of information can transmit that information poorly to subsequent processing stages. Thus, the problem of robust information propagation -- which has largely been overlooked in previous research -- may be critical for determining how our sensory organs communicate with our brains. We identify the conditions under which information propagates well -- or poorly -- through multiple stages of neural processing.

\section{Introduction}


Neurons in sensory systems gather information about the environment, and transmit that information to other parts of the nervous system. This information is encoded in the activity of neural populations, and that activity is variable: repeated presentations of the same stimulus lead to different neuronal responses~\cite{Britten93,Sof+93,faisal08,church10,franke15,zyl15,zhs15}. This variability can degrade the ability of neural populations to encode information about stimuli, leading to the question: which features of population codes help to combat -- or exacerbate -- information loss? 

This question is typically addressed by assessing the amount of information that is encoded in the periphery as a function of the covariance structure \cite{Zohary:1994ei,Abbott:1999ul,Sompolinsky:2001hh,romo03,Averbeck:2006ew,shamir06,Averbeck:2006vj,Josic:2009du,Ecker:2011bx,Cohen:2011eh,Roudi_Latham,daSilveira:2013vf,hu14,shamir14,moreno14,JZ_HOC,acg15,zyl15}, the shapes of the tuning curves~\cite{pou99,Zhang1999}, or both \cite{wilke02,Tkacik:2010hv}.
However, the informativeness of the population responses at the periphery is not the only relevant quantity for understanding sensory coding; of potentially equal importance is the amount of information that propagates through the neural circuit to downstream structures~\cite{Beck:2011cb,toyoizumi06}.

\begin{figure}[t!]
\begin{center}
\includegraphics[width=5in]{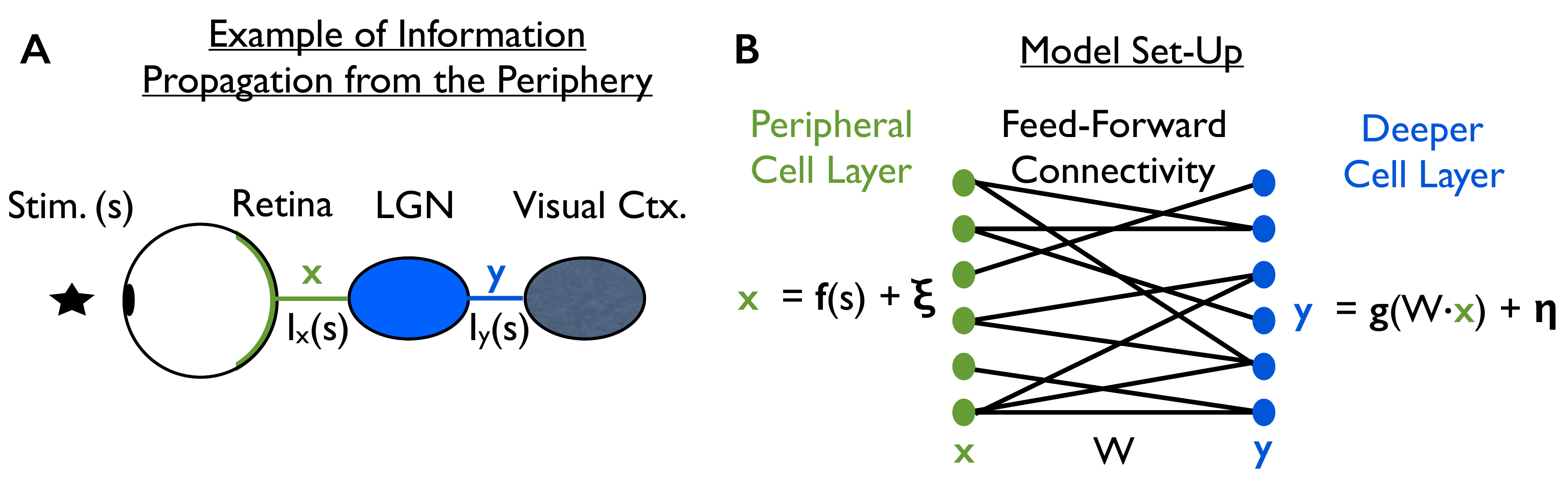}
\caption{ \textbf{The information propagation problem.} This problem is illustrated with the visual periphery, but the information propagation problem is general: it arises whenever information is transmitted from one area to another, and also when information is combined to carry out computations. (\textbf{A}) The retina transmits information about visual stimuli, $s$, to the visual cortex. The information does not propagate directly from retina to cortex; it is transmitted via an intermediary structure, the lateral geniculate nucleus (LGN). Consequently, the information about the stimuli that is available to the cortex, denoted $I_y(s)$, is not the same as the information that retina transmits, denoted $I_x(s)$. Here, we ask what properties of neural activities in the periphery maximize the information that propagates to the deeper neural structures. (\textbf{B}) Illustration of our model. Neural activity in the periphery, \textbf{x}, is generated by passing the stimulus, $s$, through a set of neural tuning curves, $\textbf{f}(s)$, and then adding zero-mean noise, {$\xxi$}, which may be correlated between cells. This activity then propagates via feed-forward connectivity, described by the matrix $\ww$, to the next layer. The activity at the next layer, \textbf{y}, is generated by passing the inputs, $\ww \cdot \xx$, through a nonlinearity $g(\cdot)$, and then adding zero-mean noise, $\eeta$. }
\label{fig:retina} 
\end{center}
\end{figure}

To illustrate the ideas, consider the case of retinal ganglion cells transmitting information about visual stimuli to the cortex via the thalamus, as shown in Fig.~\ref{fig:retina}. To quantify the performance of the retina, one must consider not only the informativeness of the optic nerve responses ($I_x(s)$ in Fig.~\ref{fig:retina}A), but also how much of that information is transmitted by the lateral geniculate nucleus (LGN) to the cortex ($I_y(s)$ in Fig.~\ref{fig:retina}A)~\cite{bejj11}. The two may be very different, as only information that survives the LGN's spike-generating nonlinearity and noise corruption will propagate to downstream cortical structures.

Despite its importance, the ability of information to propagate through neural circuits remains relatively unexplored~\cite{bejj11}. One notable exception is the literature on how synchrony among the spikes of different cells affects responses in downstream populations~\cite{Salinas:2000wr,salinas_sejno_2001,reid01,bruno11,abeles82}. This is, however, distinct from the information propagation question we consider here, as there is no guarantee that those downstream spikes will be informative.
Other work~\cite{pou99,Series:2004bw,ren11,Beck:2011cb,toyoizumi06} investigated the question of optimal network properties (tuning curves and connection matrices) for information propagation in the presence of noise.


No prior work, however, has isolated the impact of correlations on the ability of population-coded information to propagate. Given the frequent observations of correlations in the sensory periphery~\cite{Cohen:2011eh,zyl15,lampl99,alonso96,Zohary:1994ei,goris14,smith08,ecker14,scholvinck15,lin15}, and the importance of the information propagation problem, this is a significant gap in our knowledge.
To fill that gap, we consider a model (Fig.~\ref{fig:retina}B; described in more detail below), in which there are two layers (retina and LGN, for example). The first layer contains a fixed amount of information, $I_x(s)$, which is encoded in the noisy, stimulus-dependent responses of the cells in that layer. The information is passed to the second layer via feedforward connections followed by a nonlinearity, with noise added along the way. We ask how the covariance structure of the trial-to-trial variability in the first layer affects the amount of information in the second.

Although we focus on information propagation, the problem we consider  applies to more general scenarios.
In essence, we are asking: how does the noise in the input to a network interact with noise added to the output?
Because we consider linear feedforward weights followed by a nonlinearity, the possible transformations from input to output, and thus the computations the network could perform, is quite broad~\cite{hornik89}. 
Thus, the conclusions we draw apply not just to information propagation, but also to many computations.
Moreover, it may be possible to extend our analysis to recurrent, time-dependent neural networks. That is, however, beyond the scope of this work.

Our results indicate that the amount of information that successfully propagates to the second layer depends strongly on the structure of correlated responses in the first. For linear neural gain functions, and some classes of nonlinear ones, we identify analytically the covariance structures that optimize information propagation through noisy downstream circuits.
Within the optimal family of covariance structures, we find variability with so-called differential correlations \cite{moreno14} -- correlations that are proven to minimize the information in neural population activity. Thus, covariance structures that maximize the information content of neural population codes, and those that maximize the ability of this information to propagate, can be very different. Importantly, we also find that redundancy is neither necessary nor sufficient for the population code to be robust against corruption by noise.
Consequently, to understand how correlated neural activity affects the function of neural systems, we must not only consider the impact of those correlations on information, but also the ability of the encoded information to propagate robustly through multi-layer circuits.

\section{Results}

\subsection{Problem Formulation: Information Propagation in the Presence of Corrupting Noise}
\label{sec:setup}

We consider a  model in which a vector of ``peripheral" neural population responses, $\xx$, is determined by two components.  The first is the set of tuning curves, $\ff(s)$, which define the cells' mean responses to any particular stimulus (typical tuning curves are shown in Fig.~\ref{fig:redundant}A). Here we consider a one dimensional stimulus, denoted $s$, which may represent, for example, the direction of motion of a visual object. In that case, a natural interpretation of our model is that it describes the transmission of motion information by direction selective retinal ganglion cells to the visual cortex (Fig.~\ref{fig:retina})~\cite{barlow65,zyl15,franke15}. Extension to multi-dimensional stimuli is straightforward. The second component of the neural population responses, $\xxi$, represents the trial-to-trial variability. This results in the usual ``tuning curve plus noise'' model,

\begin{equation}
\label{eq:noisyX}
\xx = \ff(s) + \xxi,
\end{equation}

\noindent
where $\xxi$ is a zero mean random variable with covariance $\Sig_\xi$.


The neural activity, $\xx$, propagates to the second layer via feed-forward weights, $\ww$, as in the model of~\cite{ren11}. The activity in the second layer is given by passing the input, $\ww \cdot \xx$, through a nonlinearity, $g(\cdot)$, and then corrupting it with noise, $\eeta$ (Fig.~\ref{fig:retina}B),

\begin{equation}
\yy = g(\ww \cdot \xx) + \eeta,
\label{eq:noisyY}
\end{equation}

\noindent
where the nonlinearity is taken component by component, and $\eeta$ is zero mean noise with covariance matrix $\Sig_\eta$.
The function $g(\cdot)$ need not be invertible, so this model can include spike generation. 

While we have, in Fig.~\ref{fig:retina}, given one explicit interpretation of our model, the model itself is quite general. This means that our results apply more broadly than just to circuits in the peripheral visual system. Moreover, while our analysis (below) focuses on information loss between layers, this should not be taken to mean that there is no meaningful computation happening within the circuit: because we have considered arbitrary nonlinear transformations between layers, the same model can describe a wide range of possible computations~\cite{hornik89}. Our results apply to information loss during those computations.

In the standard fashion~\cite{moreno14,Averbeck:2006ew,hu14,shamir14,zyl15}, we quantify the information in the neural responses using the linear Fisher information. This measure quantifies the precision (inverse of the mean squared error) with which a locally optimal linear estimator can recover the stimulus from the neural responses~\cite{cramer46,rao45}. The linear Fisher information in the first and second layers, denoted $I_x(s)$ and $I_y(s)$, respectively, is given by

\begin{subequations}
\begin{align}
\label{eq:LFI_x}
I_x(s) &= \ff'(s) \cdot \Sigx^{-1} \cdot \ff'(s)
\\
\label{eq:LFI_y}
I_y(s) &= \ff'(s) \cdot
\big[
\Sig_\xi + (\weff^T \cdot \Sigeff^{-1} \cdot \weff)^{-1}
\big]^{-1} \cdot \ff'(s)
\end{align}
\end{subequations}
where a prime denotes a derivative.
Here $\weff$ are the effective weights -- basically, the weights, $\ww$, multiplied by the average slope of the gain function, $g(\cdot)$ -- and $\Sigeff$ includes contributions from the noise in the second layer, $\eeta$, and, if $g(\cdot)$ is nonlinear, from the noise in the first layer. (If $g$ is linear, $\Sigeff = \Sig_\eta$, so in this case $\Sigeff$ depends only on the noise in the second layer). This expression is valid if $\weff^T \cdot \Sigeff^{-1} \cdot \weff$ is invertible; so long as there are more cells in the second layer than the first, this is typically the case. See Methods, Sec.~\ref{sec:iout}, for details.

Equation~\eqref{eq:LFI_y} is somewhat intuitive, at least at a gross level: both large effective noise ($\Sigeff$) and small effective weights ($\weff$) reduce the amount of information at the second layer. At a finer level, the relationship between the two covariance structures -- corresponding to the first and second terms in brackets in Eq.~\eqref{eq:LFI_y} -- can have a large effect on $I_y(s)$, as we will see shortly.

\subsection{Information content and information propagation put different constraints on neural population codes}
\label{sec:propa_vs_rep}

\begin{figure}[t!]
\begin{center}
\includegraphics[width=6in]{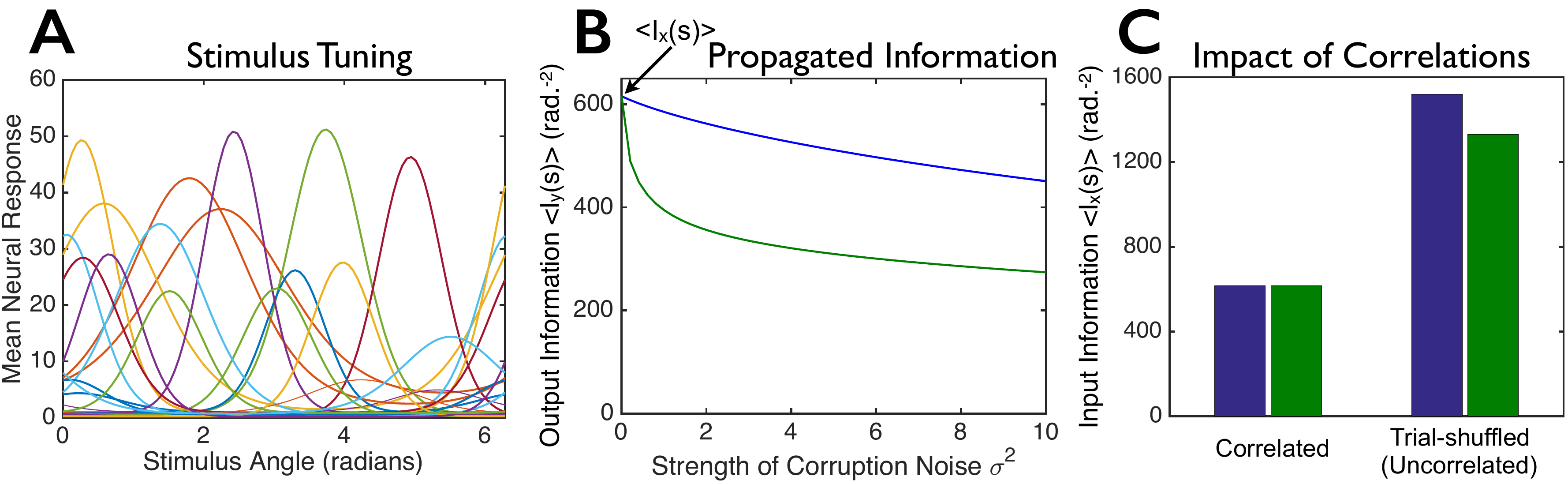}
\caption{ \textbf{Not all population codes are equally robust against corruption by noise.} We constructed two model populations, each with the same 100 tuning curves for the first layer of cells but with different covariance structures, $\Sigx$ (see text, especially Eq.~\eqref{eq:covs_green_blue}). The covariance structures were chosen so that the two populations convey identical amounts of information $I_x(s)$ about the stimulus.
(\textbf{A}) 20 randomly-chosen tuning curves from the 100 cell population.
(\textbf{B}) We corrupted the responses of each neural population by additional Gaussian noise (independently and identically distributed for all cells) of variance $\sigma^2$, to mimic corruption that might arise as the signals propagate through a multi-layered neural circuit, and computed the ``output" information $I_y(s)$ that these further-corrupted responses convey about the stimulus (blue and green curves). The population shown in green forms a relatively fragile code wherein modest amounts of noise strongly reduce the information, whereas the population shown in blue is more robust. (\textbf{C}) Input information $I_x(s)$ in the two model populations (left; ``correlated") and information that would be conveyed by the model populations if they had their same tuning curves and levels of trial-to-trial variability, but no correlations between cells (right; ``trial-shuffled"). For panels B and C, we computed the information for each of 100 equally spaced stimulus values, and averaged the information over those stimuli. See Methods, Sec.~\ref{sec:fig2} for additional details.}
\label{fig:redundant}
\end{center}
\end{figure}

We begin with an example to highlight the difference between the information contained in neural population codes and the information that propagates through subsequent layers. Here, we consider two different  neuronal populations with identical tuning curves (Fig.~\ref{fig:redundant}A), nearly-identical levels of trial-to-trial neural variability, and identical amounts of stimulus information encoded in their firing-rate responses; the populations' correlational structures, however, differ. We then corrupt these two populations' response patterns with noise, to mimic corruption that might arise in subsequent processing stages, and ask how much of the stimulus information remains. Surprisingly, the two population codes can show very different amounts of information after corruption by even modest amounts of noise (Fig.~\ref{fig:redundant}B).

In more detail, there are 100 neurons in the first layer; those neurons encode an angle, denoted $s$, via their randomly-shaped and located tuning curves (Fig.~\ref{fig:redundant}A). We consider two separate model populations. Both have the same tuning curves, but different covariance matrices. For reasons we discuss below, those covariance matrices, denoted $\Sigx^{\text{blue}}$ and $\Sigx^{\text{green}}$ (blue and green correspond to the colors in Figs.~\ref{fig:redundant}B and C), are given by

\begin{subequations}
\label{eq:covs_green_blue}
\begin{align}
\Sigx^{\text{blue}} &= \Sig_0 + \epsilon \bf{f'}(s) \bf{f'}(s)
\\
\Sigx^{\text{green}} &= \Sig_0  + \epsilon_u \uu(s) \uu(s) 
\end{align}
\end{subequations}
where $\Sig_0$ is a diagonal matrix with elements equal to the mean response,
\begin{align}
\label{sig0}
\Sigma_{0,ij} = f_i(s) \delta_{ij} .
\end{align}
Here $\delta_{ij}$ is the Kronecker delta ($\delta_{ij} = 1$ if $i=j$ and 0 otherwise), and we use the convention that two adjacent vectors denote an outer product; for instance, the $ij^{\rm th}$ element if $\uu \uu$ is $u_i u_j$.
The vector $\uu$ has the same magnitude as $\ff'$, but points in a slightly different direction (it makes an angle $\theta_u$ with $\fprime$), and $\epsilon$ and $\epsilon_u$ are chosen so that the information in the two populations, $I_x(s)$, is the same ($\epsilon_u$ also depends on $s$; we suppress that dependence for clarity).

In our simulations, both $\epsilon$ and $\epsilon_u$ are small (on the order of $10^{-3}$; see Sec.~\ref{sec:fig2}), so the variance of the $i^{\rm th}$ neuron is approximately equal to its mean. This makes the variability Poisson-like, as is typically observed when counting neural spikes in finite time windows~\cite{Britten93,Sof+93,faisal08,zyl15,church10,franke15}. (More precisely, the average Fano factors -- averaged over neurons and stimuli -- were 1.01 for the ``blue" population and 1.04 for the ``green" one.)  Both model populations also have the same average correlation coefficients, which are near-zero (see Methods, Sec.~\ref{sec:fig2}, for details).

To determine how much of the information in the two populations propagates to the second layer, we computed $I_y(s)$ for both populations using Eq.~\eqref{eq:LFI_y}.
For simplicity, we used the identity matrix for the feed-forward weights, $\ww$, a linear gain function, $g(\cdot)$, and independently and identically distributed (\textit{iid}) noise with variance $\sigma^2$. Later we consider the more general case: arbitrary feedforward weights, nonlinear gain functions, and arbitrary covariance for the second layer noise. Those complications don't, however, change the basic story.

Figure~\ref{fig:redundant}B shows the information in the output layer versus the level of output noise, $\sigma^2$, for the two populations. Blue and green curves correspond to the different covariance structures. Although the two populations have identical tuning curves, nearly-identical levels of trial-to-trial neural variability, and contain identical amounts of information about the stimulus, they differ markedly in the robustness of that information to corruption by noise in the second layer. Thus, quantifying the information content of neural population codes is not sufficient to characterize them: recordings from the first-layer cells of the two example populations in Fig.~\ref{fig:redundant} would yield identical information about the stimulus, but the blue population has a greater ability to propagate that information downstream. 

One possible explanation for the difference in robustness is that the information in the green population relies heavily on correlations, which are destroyed by a small amount of noise. To check this, we compared the information of the correlated neural populations to the information that would be obtained with the same tuning curves and levels of single neuron trial-to-trial variability, but no inter-neuronal correlations~\cite{Schneidman:2003ej,romo03,adibi13} (Fig.~\ref{fig:redundant}C). We find that removing the correlations actually {\it increases} the information in both populations (Fig.~\ref{fig:redundant}C; ``Trial-Shuffled"), and by about the same amount, so this possible explanation cannot account for the difference in robustness. We also considered the case where the correlated responses carry more information than would be obtained from independent cells. We again found (similar to Fig. 2) that there could be substantial differences in the amount of information propagated by equally informative population codes (see Methods, Fig.~\ref{fig:synergy}).

These examples illustrate that merely knowing the amount of information in a population, or how that information depends on correlations in neural responses, doesn't tell us how much of that information will propagate to the next layer. In the remainder of this paper, we provide a theoretical explanation of this observation, and identify the covariance structures at the first layer that maximize robustness to information loss during propagation through downstream circuits.

\subsection{Geometry of Robust Versus Fragile Population Codes}
\label{sec:robust_vs_fragile}

\begin{figure}[t!]
\begin{center}
\includegraphics[width=3.0in]{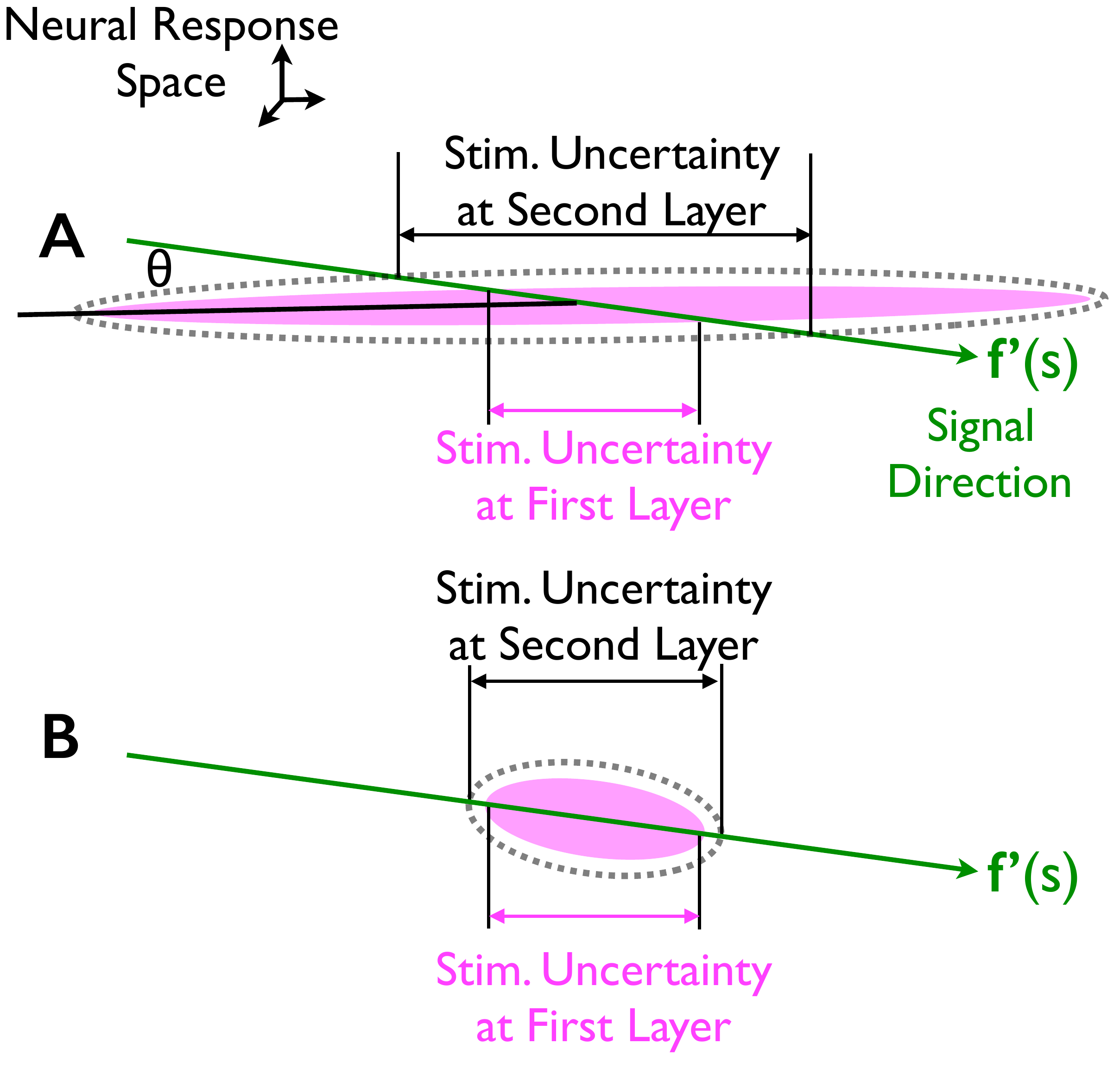}
\caption{ \textbf{Geometry of robust versus fragile population codes.} Cartoons showing the interaction of signal and noise for two populations with the same information in the input layer. The dimension of the space is equal to the number of cells in the population; we show a two dimensional projection.
Within this space, when the stimulus changes by an amount $\Delta s$ (with $\Delta s$ small), the average neural response changes by $\fprime(s) \Delta s$. Thus, $\fprime(s)$ is the ``signal direction" (green arrows). Trial-by-trial fluctuations in the neural responses in the first layer are described by the ellipses; these correspond to 1 standard-deviation probability contours of the conditional response distributions. The impact of the neural variability on the encoding of stimulus $s$ is determined by the projection of the response distributions onto the signal direction (magenta double-headed arrows). By construction, these are identical in the first layer. Accordingly, an observer of the neural activity in the first layer of either population would have the same level of uncertainty about the stimulus, and so both populations encode the same amount of stimulus information. When additional \textit{iid} noise is added to the neural responses, the response distributions grow; the dashed ellipses show the resultant response distributions at the second layer. Even though the same amount of \textit{iid} noise is added to both populations, the one in panel A shows greater stimulus uncertainty after the addition of noise than does the one in panel B. Consequently, the information encoded by the population in panel B is more robust against corruption by noise.}
\label{fg:skew_cartoons} 
\end{center}
\end{figure}

To understand, from a geometrical point of view, why some population codes are more sensitive to noise than others, we need to consider the relationship between the noise covariance ellipse and the ``signal direction," $\fprime(s)$ -- the direction the mean neural response changes when the stimulus $s$ changes by a small amount.
Figures~\ref{fg:skew_cartoons}A and B show this relationship for two different populations. The noise distribution in the first layer is indicated by the magenta ellipses, and the signal direction by the green arrows.
The uncertainty in the stimulus after observing the neural response is indicated by the overlap of the green line with the magenta ellipse.
Because the overlap is the same for the two populations, they have the same amount of stimulus uncertainty, and thus the same amount of information -- at least in the first layer.

Although the two populations have the same amount of information, the covariance ellipses are very different: one long and skinny but slightly tilted relative to the signal direction (Fig.~\ref{fg:skew_cartoons}A), the other shorter and fatter and parallel to the signal direction (Fig.~\ref{fg:skew_cartoons}B).
Consequently, when \textit{iid} noise is added, as indicated by the dashed lines, stimulus uncertainty increases by very different amounts:
there's a much larger increase for the long skinny ellipse than for the short fat one.
This makes the population code in Fig.~\ref{fg:skew_cartoons}A much more sensitive to added noise than the one in Fig.~\ref{fg:skew_cartoons}B.

To more rigorously support this intuition, in Methods, Sec.~\ref{sec:info_loss}, we derive explicit expressions for the stimulus uncertainty in the first and second layers as a function of the angle between the long axis of the covariance ellipse and the signal direction. Those expressions corroborate the phenomenon shown in Fig.~\ref{fg:skew_cartoons}.

\subsection{A family of optimal noise structures}
\label{sec:gen_dist}

The geometrical picture in the previous section tells us that a code is robust against added noise if the covariance ellipse lines up with the signal direction.Taken to it's extreme, this suggests that when all the noise is concentrated along the $\fprime(s)$ direction, so that the covariance matrix is given by
\begin{equation} 
\label{eq:optsigxi}
\Sigx(s) \propto \fprime(s)\fprime(s) ,
\end{equation}
the resulting code should be optimally robust.
While this may be intuitively appealing, the arguments that led to it were based on several assumptions: \textit{iid} noise added in the second layer, feedforward weights, $\ww$, set to the identity matrix, and a linear neural response function $g(\cdot)$. In real neural circuits, none of these assumptions hold. It turns out, though, that the only one that matters is the linearity of $g(\cdot)$. In this section we demonstrate that the covariance matrix given by Eq.~\eqref{eq:optsigxi} optimizes information transmission for neurons with linear gain functions (although we find, perhaps surprisingly, that this optimum is not unique). In the next section we consider nonlinear gain functions; for that case the covariance matrix given by Eq.~\eqref{eq:optsigxi} can be, but is not always guaranteed to be, optimal.

To determine what covariance structures maximize information propagation, we simply maximize information in the second layer, $I_y(s)$, with respect to the noise covariance matrix in the first layer, $\Sig_\xi$, with the information in the first layer held fixed. When the gain function, $g(\cdot)$, is linear (the focus of this section), this is relatively straightforward. Details of the calculation are given in Methods, Sec.~\ref{sec:identify_optimal_family}; here we summarize the results.

The main finding is that there exists a family of first-layer covariance matrices $\Sigx$, not just one, that maximizes the information in the second layer. That family, parameterized by $\alpha$, is given by 
\begin{eqnarray}
\label{optimal_noise}
\Sigx(s) = \frac{\alpha}{I_x(s)} \, \ieta(s) \Sigy +
\frac{1-\alpha}{I_x(s)} \,
\fprime(s) \fprime(s),
\end{eqnarray} 
where $\Sig_y$ is the effective covariance matrix in the second layer, 

\begin{align}
\label{sigydef}
\Sigy \equiv (\weff^T \cdot \Sige^{-1} \cdot \weff)^{-1}
,
\end{align}
and $I_\eta(s)$ is the information the second layer would have if there were no noise in the first layer,
\begin{eqnarray}
\label{ieta}
\ieta(s) =  \fprime(s) \cdot \Sigy^{-1} \cdot \fprime(s) 
\end{eqnarray} 
(see in particular Methods, Eq.~\eqref{general_optimal_noise}).
For this whole family of distributions -- that is, for any value of $\alpha$ for which $\Sigx$ is positive semi-definite -- the output information, $I_y(s)$, has exactly the same value,
\begin{equation}
I_y(s) = \frac{I_x(s)}{1 + I_x(s)/I_\eta(s)}
\end{equation}
(see Methods, Eq.~\eqref{info_ratio}). This is the maximum possible output information given the input information, $I_x(s)$. 

\begin{figure}[t!]
\begin{center}
\includegraphics[width=4in]{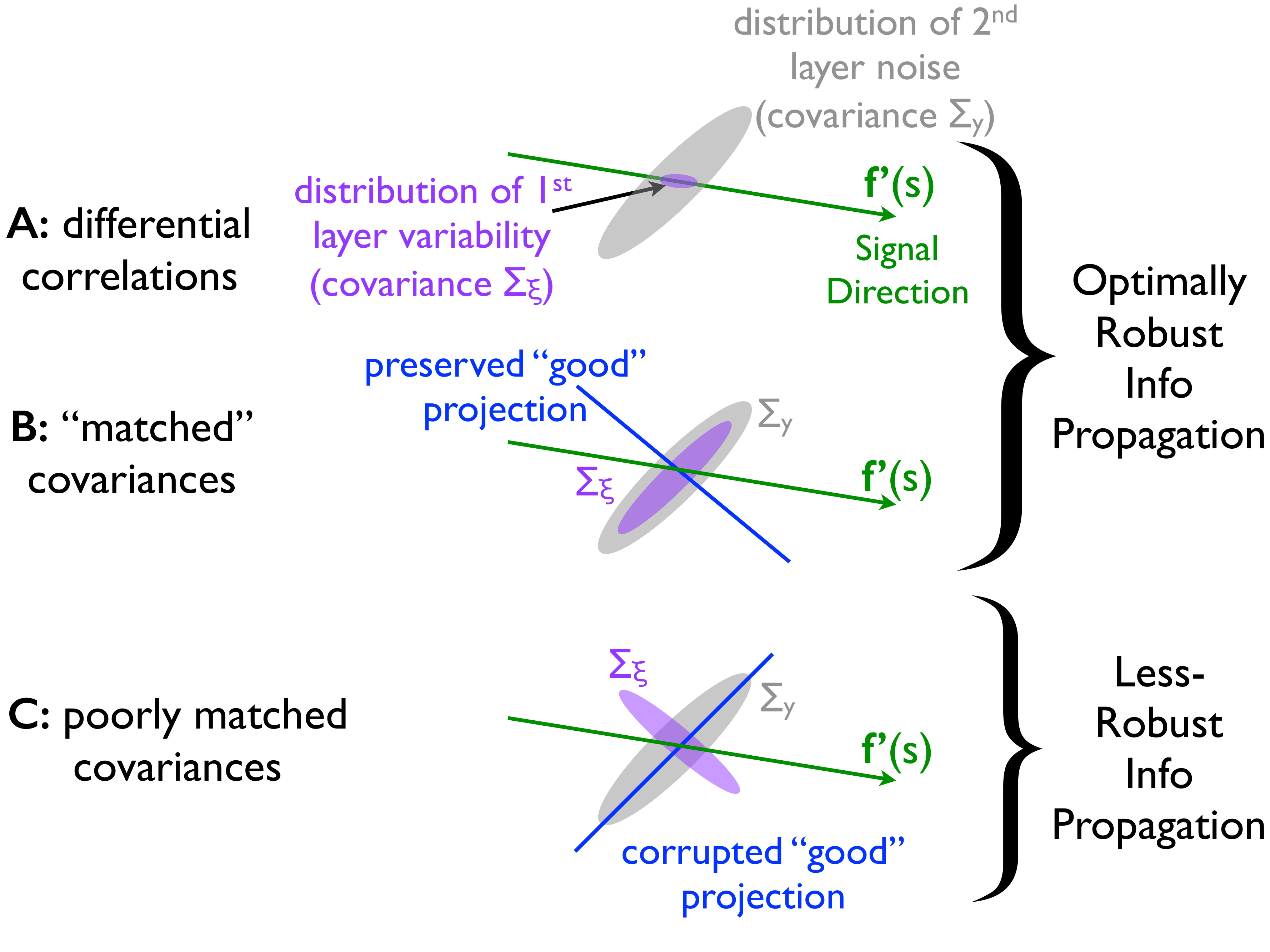}
\caption{\textbf{Family of optimal covariance matrices.} For all panels, green arrows indicate the signal direction, $\fprime(s)$. Magenta ellipses indicate the noise in the first layer (with corresponding covariance matrix $\Sigx$), and grey ellipses indicate the effective noise in the second layer (with corresponding covariance matrix $\Sigy$).
(\textbf{A}) The covariance ellipse in the first layer has its long axis aligned with the signal direction; this configuration (which corresponds to differential correlations) optimizes information robustness for any distribution of second layer noise.
(\textbf{B})  The covariance ellipse in the first layer does not have its long axis aligned with the signal direction. However, the covariance ellipse of the effective noise in the second layer, $\Sigy$, has the same shape as the covariance ellipse in the first. In this case, the blue ``good" projection  -- which is aligned both with a low-variance direction of the first-layer distribution (magenta), and with the signal curve (green), and thus is relatively informative about the stimulus (see text) -- is corrupted by relatively little noise at the second layer. This ``matched" noise configuration is among those that optimize robustness to noise. The optimal family of covariance matrices interpolates between the configurations shown in panels A and B.
(\textbf{C}) Again the covariance ellipse in the first layer does not have its long axis aligned with the signal direction. But now the  ``good" projection is heavily corrupted by noise at the second layer. In this configuration, all projections are substantially corrupted by noise at some point in the circuit, and thus relatively little information can propagate.}
\label{fg:matching}
\end{center}
\end{figure}

Two members of this family are of particular interest. One is $\alpha=0$, for which the covariance matrix corresponds to differential correlations (Eq.~\eqref{eq:optsigxi}); that covariance matrix is illustrated in Fig.~\ref{fg:matching}A. This covariance matrix aligns the noise direction with the signal direction. Accordingly, as for the geometrical picture in Fig.~\ref{fg:skew_cartoons}, it makes the encoded information maximally robust.

The other family member we highlight is $\alpha=1$, for which $\Sigx \propto \Sigy$. For this case, the covariance matrix in the first layer matches the effective covariance matrix in the second layer; we thus refer to this as ``matched covariance". To understand why this covariance optimizes information in the second layer, we start with the observation that the population activities can be decomposed into their principal components: each principal component corresponds to a different axis along with the population activities can be projected. The information contained in each such projection (principal component) adds up to give the total Fisher information (see Methods, Eq.~\ref{info_models_sum}). The most informative of these projections are those that have low noise variance, and which align somewhat with the signal curve -- like the blue line in Fig.~\ref{fg:matching}B. When $\Sigx \propto \Sigy$, the projections that are most informative in the first layer are corrupted by relatively little noise in the second layer. Consequently, this configuration enables robust information propagation.
In contrast, when the covariance structures in the first and second layers are less well matched, all projections are heavily corrupted by noise at some point (i.e., either in the first or the second layer), and hence very little information propagates (Fig.~\ref{fg:matching}C). 

The family of optima interpolates between the two configurations shown in Figs.~\ref{fg:matching}A and B (see also Eq.~\eqref{optimal_noise}). Almost all members of this optimal covariance family depend on the details of the downstream circuit: for $\alpha \ne 0$ in Eq.~\eqref{optimal_noise}, the optimal noise covariance at the first layer depends on the feed-forward weights, $\ww$, and the structure of the downstream noise. The one exception to this is the covariance matrix given by Eq.~\eqref{eq:optsigxi}: that one is optimal regardless of the downstream circuit. These are so-called ``differential correlations'' -- the only correlations that lead to information saturation in large populations \cite{moreno14}, and the correlations that minimize information in general (Methods, Sec.~\ref{sec:mininfo}). The fact that correlations can minimize information content and at the same time maximize robustness highlights the fact that optimizing the amount of information in a population code versus optimizing the ability of that information to be transmitted put very different constraints on neural population codes.

The existence of an optimum where the covariance matrices are matched across layers emphasizes that not all optimally robust population codes are necessarily redundant. (By redundant we mean the population encodes less information than would be encoded by a population of independent cells with the same tuning curves and levels of single neuron trial-to-trial variability \cite{Averbeck:2006ew,shamir14}; see Fig.~\ref{fig:redundant}). Notably, if the effective second layer covariance matrix, $\Sigy$, admits a synergistic population code -- wherein more information is encoded in the correlated population versus an uncorrelated one with the same tuning curves and levels of trial-to-trial response variability -- then the matched case, $\Sigx \propto \Sigy$, will also admit a synergistic population code, and be optimally robust.

Optimally robust, however, does not necessarily mean the majority of the information is transmitted; for that we need another condition.
We show in Sec.~\ref{sec:variances_strategies} that for non-redundant codes, a large fraction of the information is transmitted only if there are many more neurons in the second layer than in the first.
This is typically the case in the periphery.
For differential correlations, that condition is not necessary -- so long as there are a large number of neurons in both the input and output layers, most of the information is transmitted. 

\subsection{Nonlinear Gain Functions}
\label{sec:DG_spikes}

So far we have focused on linear gain functions $g(\cdot)$; here we consider nonlinear ones.
This case is much harder to analyze, as the effective covariance structure in the second layer, $\Sigeff$, depends on the noise in the first layer (see Methods, Sec.~\ref{sec:iout}, especially Eq.~\eqref{Sigeff_def}).
We therefore leave the analysis to Methods (Sec.~\ref{sec:opt_cov_nonlin}); here we briefly summarize the main results. After that we consider two examples of nonlinear gain functions -- both involving a thresholding nonlinearity to mimic spike generation.

For linear gain functions we were able to find a whole family of optimal covariance structures, for nonlinear ones we did not even try. Instead, we asked: under what circumstances are differential correlations optimal?
Even for this simplified question a definitive answer does not appear to exist.
Nevertheless, we can make progress in special cases. When there is no added noise in the second layer (e.g., $\eeta=0$ for the model in Fig.~\ref{fig:retina}B), differential correlations maximize the amount of information that propagates through the nonlinearity, so long as the tuning curves are sufficiently dense relative to the steepness of the tuning curves (meaning that whenever the stimulus changes, the average stimulus-evoked response of at least one neuron also changes; see Methods, Sec.~\ref{sec:opt_cov_nonlin}).
If there is added noise at the second layer, differential correlations tend to be optimal in cases where the addition of noise at the first layer, $\xi$, causes reductions in information, $I_x(s)$. (This means that, so long as there are no \emph{stochastic resonance} effects causing added noise to increase information, then differential correlations are optimal.)

We first check, with simulations, the prediction that differential correlations are optimal if there is no added noise. For that we use a thresholding nonlinearity, chosen for two reasons: it is an extreme nonlinearity, and so should be a strong test of our theory, and it is somewhat realistic in that it mimics spike generation. For this model, the responses at that second layer, $y_i$, are given by

\begin{align}
\label{spike}
y_i = \Theta(x_i - \theta_i)
\end{align}
where $\Theta$ is the Heaviside step function ($\Theta(x) = 1$ if $x \ge 0$ and 0 otherwise), and $\theta_i$ is the spiking threshold of the $i^{th}$ neuron.
This is the popular dichotomized Gaussian model~\cite{Macke:2009dx,Macke:2011gw,yu11jns,Amari:2003ul,Bethge:tv}, which has been shown to provide a good description of population responses in visual cortex, at least in short time windows~\cite{yu11jns}, and to provide high-fidelity descriptions of the responses of integrate-and-fire neurons, again in short time windows~\cite{leen13}.  

In our simulations with the step function nonlinearity, as for all of the other cases we considered above, the first layer responses are given by the tuning curve plus noise model (Eq.~\eqref{eq:noisyX}).
The tuning curves, $\ff(s)$, of the 100-neuron population are again heterogeneous (similar to those in Fig.~\ref{fig:redundant}A but with a different random draw from the tuning curve distribution), and the trial-to-trial variability is given by
\begin{align}
\label{sigma_uu}
\Sigx = \gamma_u
\big[ \Sig_0 + \epsilon_u \uu(s) \uu(s) \big]
\end{align}
with $\Sig_0$ given by Eq.~\eqref{sig0}.
This is the same covariance matrix as in Eq.~(\ref{eq:covs_green_blue}b), except that we have included an overall scale factor, $\gamma_u$, chosen to ensure that the information in the input layer is independent of both $\epsilon_u$ and $\uu(s)$ (see Methods, Sec.~\ref{sec:info_uu}, especially Eq.~\eqref{eq:gamma_u}). 

\begin{figure}[t!]
\begin{center}
\includegraphics[width=5in]{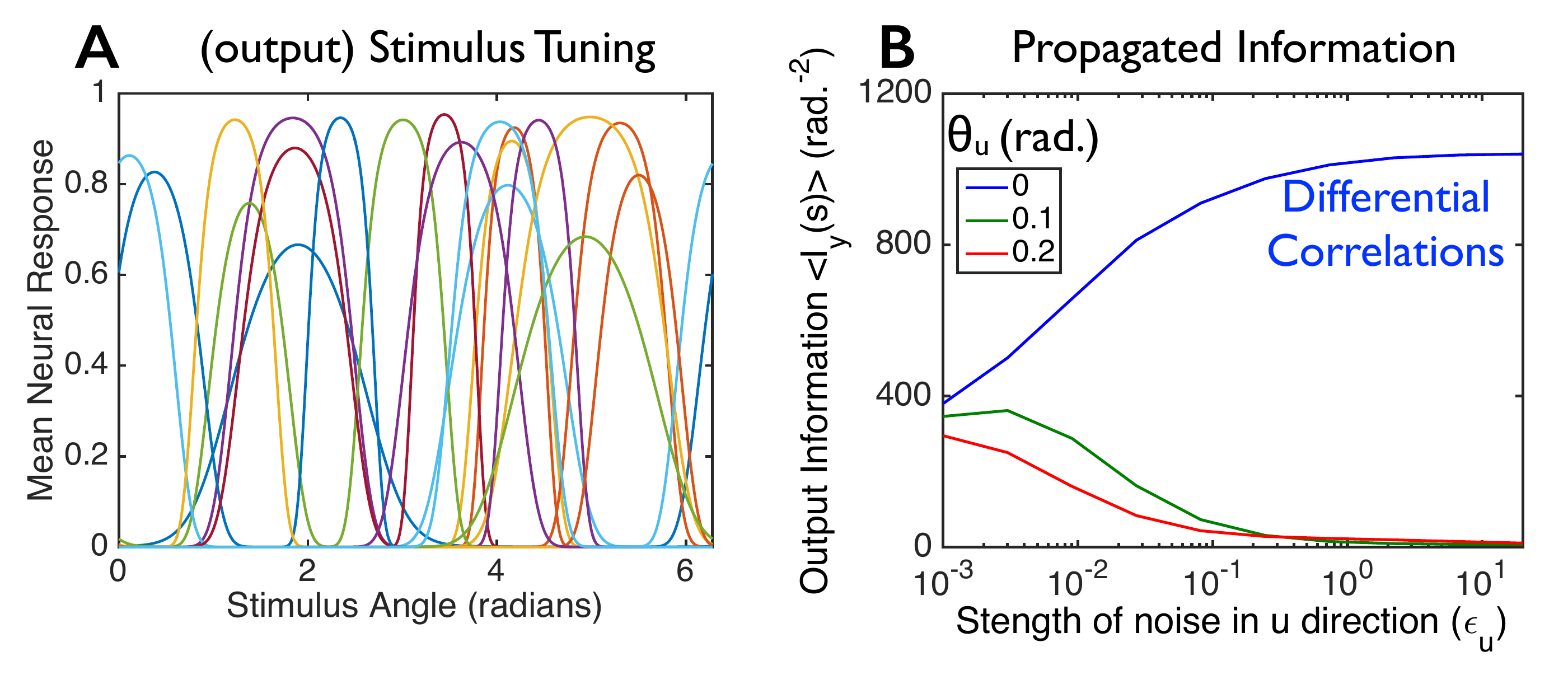}
\caption{ \textbf{Differential correlations enhance information propagation through ``spike-generating" nonlinearities.} Responses in the second layer were generated using the dichotomized Gaussian model of spike generation, in which the input from the first layer was simply binarized via a step function (see Eq.~\eqref{spike}). We varied the correlations in these inputs (see Eq.~\eqref{sigma_uu}) while keeping the input information and input tuning curves fixed. (\textbf{A}) Heterogeneous tuning curves in the second layer, evaluated at $\epsilon_u = 0$; we show a random subset of 20 cells out of the 100-neuron population studied in panel B. (\textbf{B}) Information transmitted by the 100-cell spiking population as a function of $\epsilon_u$, which is the strength of the noise in the $\uu(s)$ direction, for different angles, $\theta_u$, between $\uu$ and $\fprime(s)$ (see Eq.~\eqref{sigma_uu}). The input information 
was held fixed as $\epsilon_u$ was varied. The information is averaged over 20 evenly spaced stimuli (see Methods, Sec.~\ref{sec:fig5}). }
\label{fg:fig_DG}
\end{center}
\end{figure}

Because these (step function) nonlinearities are infinitely steep, the tuning curves are not sufficiently dense for our mathematical analysis to guarantee that differential correlations are optimal for information propagation.  However, we argue in Methods, Sec.~\ref{sec:opt_cov_nonlin}, that this should be approximately true for large populations. And indeed, that's what we find with our numerical simulation, as shown in Fig.~\ref{fg:fig_DG}B. When $\theta_u = 0$ (recall that $\theta_u$ is the angle between $\uu(s)$ and $\fprime(s)$), so that $\uu(s) = \fprime(s)$, the second term in Eq.~\eqref{sigma_uu} corresponds to differential correlations; in this case, information increases monotonically with $\epsilon_u$. In other words, information propagated through the step function nonlinearity increases as ``upstream" correlations become more like pure differential correlations.
In contrast, when $\theta_u$ is nonzero (as in Fig.~\ref{fg:skew_cartoons}A), information does not propagate well: information decreases as $\epsilon_u$ increases. This is consistent with our findings for the linear gain function considered in Fig.~\ref{fig:redundant}. Thus, differential correlations can optimize information transmission even for a nonlinearity as extreme as a step function.

The lack of explicit added noise at the second layer makes this case somewhat unrealistic. In neural circuits, we expect noise to be added at each stage of processing -- if nothing else, due to synaptic failures. We thus considered a model in which noise is added before the spike-generation process,

\begin{align}
\label{spike_noise}
y_i = \Theta(x_i + \zeta_i - \theta_i)
\end{align}
where $\zeta_i$ is zero-mean noise with covariance matrix $\Sig_\zeta$.

\begin{figure}[t!]
\begin{center}
\includegraphics[width=5in]{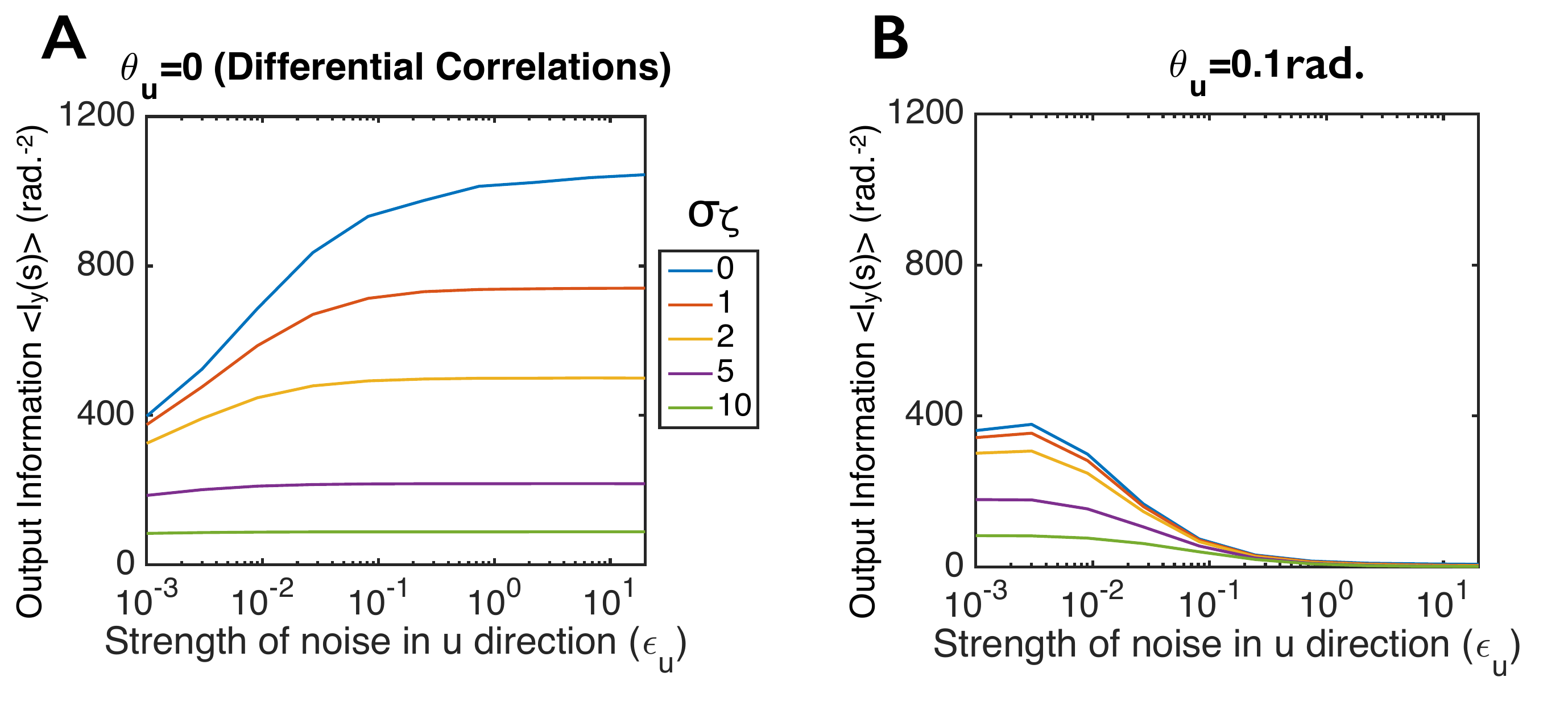}
\caption{ \textbf{Information propagation through spike-generating nonlinearities with additive input noise.} As with Fig.~\ref{fg:fig_DG}, responses in the second layer were generated using the dichotomized model of spike generation, in which the input from the first layer was simply binarized. Here, though, Gaussian noise was added before thresholding; see Eq.~\eqref{spike_noise}. We varied the correlations in the input layer (see Eq.~\eqref{sigma_uu}) while keeping the input information and input tuning curves fixed for the 100-cell population (same tuning curves and covariance matrices as in Fig.~\ref{fg:fig_DG}). The additive noise at the second layer (the $\zeta_i$) was \textit{iid} Gaussian, with variance $\sigma^2_\zeta$; different colored lines correspond to different values of $\sigma^2_\zeta$.
(\textbf{A}) Output information versus $\epsilon_u$ for populations with differential correlations ($\uu = \fprime(s)$).
(\textbf{B}) Same as panel A, but for populations that concentrate noise along an axis, $\uu$, that makes an angle of 0.1 rad with the $\fprime(s)$ direction. %
For both panels, the input information was held fixed as $\epsilon_u$ was varied, and the information was averaged over 20 evenly spaced stimuli. }
\label{fg:fig_DG_Zeta}
\end{center}
\end{figure}

We computed information for this model using the same input tuning curves, spike thresholds, and covariance matrix, $\Sig_\xi$, as without the additional noise (i.e., as in Fig.~\ref{fg:fig_DG}). To mimic the kind of independent noise expected from synaptic failures, we chose the $\zeta_i$ to be \textit{iid}, and for simplicity we took them to be Gaussian distributed with variance $\sigma^2_\zeta$. We computed the amount of stimulus information, $I_y(s)$, for several different levels of the added input noise $\sigma^2_\zeta$. We found that for all levels of noise, differential correlations increase information transmission ($I_y(s)$ increases monotonically with $\eps_u$ in Fig.~\ref{fg:fig_DG_Zeta}A, for which $\theta_u=0$). And we again found that when the long axis of the covariance ellipse makes a small angle with the signal direction, information propagates
poorly (Fig.~\ref{fg:fig_DG_Zeta}B, for which $\theta_u=0.1$ rad.). 

These numerical findings for a spike-generating nonlinearity with added noise are similar to the previous cases of a linear transfer function, $g(\cdot)$, with added input noise (Figs.~\ref{fig:redundant} and \ref{fg:skew_cartoons}), for which we have analytical results, or 
a spike generating nonlinearity with no added input noise (Fig.~\ref{fg:fig_DG}), for which we do not. We further argue in Methods, Sec.~\ref{sec:opt_cov_nonlin}, that for nonlinear gain functions differential correlations are likely to be optimal if the tuning curves are optimal (in the case of Eq.~\eqref{spike_noise}, if the thresholds $\theta_i$ are chosen optimally). Taken together, our findings demonstrate that differential correlations in upstream populations generally increase the information that can be propagated downstream through noisy, nonlinear neural circuits.  



\section{Discussion}

Much work in systems neuroscience has investigated the factors that influence the amount of information about a stimulus that is encoded in neural population activity patterns. Here we addressed a related question that is often overlooked: how do correlations between neurons affect the ability of information to propagate robustly through subsequent stages of neural circuitry? The question of robustness is potentially quite important, as the ability of information to propagate determines how much information from the periphery will reach the deeper neural structures that affect decision making and behavior. To investigate this issue, we considered a model with two cell layers. We varied the covariance matrix of the noise in the first layer (while keeping the tuning curves and information in the first layer fixed), and asked how much information could propagate to the second layer. Our main findings were threefold.

First, population codes with different covariance structures but identical tuning curves and equal amounts of encoded information can differ substantially in their robustness to corruption by additional noise (Figs.~\ref{fig:redundant}, \ref{fg:fig_DG}, \ref{fg:fig_DG_Zeta}, and ~\ref{fig:synergy}). Consequently, measurements of information at the sensory periphery are insufficient to understand the ability of those peripheral structures to propagate information to the brain, as that propagation process inevitably adds noise. For instance, populations of independent neurons can be much worse at transmitting information than can populations displaying correlated variability (Fig.~\ref{fg:fig_DG}B). Thus, to understand how the brain efficiently encodes information, we must concern ourselves not just with the amount of information in a population code, but also with the robustness of that encoded information against corruption by noise.

Second, for linear gain functions, or noise-free nonlinear ones with sufficiently dense tuning curves, populations with so-called differential correlations \cite{moreno14} are maximally robust against noise induced by information propagation.
This fact may seem surprising given that differential correlations are the only ones that lead to information saturation in large populations~\cite{moreno14}, and the correlations that minimizme information in general (Methods, Sec.~\ref{sec:mininfo}). However, in hindsight it makes sense: differential correlations correspond to a covariance ellipse aligned with the signal direction (see Fig.~\ref{fg:skew_cartoons}B), and added noise simply doesn't make it much longer. For nonlinear gain functions combined with arbitrary noise, differential correlations are not guaranteed to yield a globally optimal population code for information propagation. However, for the spike-generating nonlinearity we considered here, differential correlations were at least a local optimum (see Figs.~\ref{fg:fig_DG} and \ref{fg:fig_DG_Zeta}).  

Third, while differential correlations optimize robustness, for linear gain functions that optimum is not unique. Instead, there is a continuous family of covariances that exhibit identical robustness to noise (see Fig.~\ref{fg:matching} and Eq.~\eqref{optimal_noise}). However, within this family, only differential correlations yield population codes that are optimally robust independent of the downstream circuitry. Thus, they are the most flexible of the optima: for all other members of the family, the optimal covariance structure in the first layer depends on the noise in subsequent layers, as well as the weights connecting those layers.

The existence of this family of optimal solutions raises an important point with regards to redundancy and robust population coding. Populations with differential correlations -- which are among the optimal solutions in terms of robustness -- are highly redundant: a population with differential correlations encodes much less information than would be expected from independent populations with the same tuning curves and levels of trial-to-trial variability (Fig.~\ref{fig:redundant}C). It is common knowledge that redundancy can enhance robustness of population codes against noise \cite{Barlow:2001ub}, and thus it is worth asking if our robust population coding results are simply an application of this fact. Importantly, the answer is no: as discussed in Sec.~\ref{sec:gen_dist},
within the family of optimal correlational structures are codes with minimal redundancy. Moreover, as is shown in Fig.~\ref{fig:redundant}B, a code can be redundant without being robust to added noise. In other words, redundancy in a population code is neither necessary, nor sufficient, to ensure that the encoded information is robust against added noise. However, there is an important caveat: unless the 
number of neurons in the second layer is large relative to the number in the first, and/or the added noise in the second layer is small relative to the noise at the first layer, non-redundant codes tend to lose a large amount of information when corrupted by noise. This contrasts sharply with differential correlations, which can tolerate large added noise with very little information loss (see Sec.~\ref{sec:variances_strategies}).

In the case of real neural systems, there will always be a finite amount of information that the population can convey (bounded by the amount of input information that the population receives from upstream sources~\cite{Beck:2012vm}), and so the question of how best to propagate a (fixed) amount of information is of potentially great relevance for neural communication. Our results suggest that the presence of differential correlations serves to allow population-coded information to propagate robustly. Thus, an observation of these correlations in neural recordings might indicate that the population code is optimized for robustness of the encoded information. At the same time, we note that weak differential correlations might be hard to observe experimentally~\cite{moreno14}. Moreover, our calculations indicate that there exists a whole family of possible propagation-enhancing correlation structures, and so differential correlations are not necessary for robust information propagation. This means that observations of either differential correlations, correlation structures matched between subsequent layers of a neural circuit (Fig. 4), or a combination of the above would indicate that the system enables robust information propagation.

How might the nervous system shape its responses so as to generate correlations that enhance information propagation? Recent work identified network mechanisms that can lead to differential correlations~\cite{kan15}. While it is beyond the scope of this work, it would be interesting to explicitly study the network structures that allow encoded information to propagate most robustly through downstream circuits.
Relatedly, \cite{ren11} and~\cite{Beck:2011cb,toyoizumi06} asked how the connectivity between layers affects the ability of information to propagate. While we identified the optimal patterns of input to the multi-stage circuit, they identified the optimal anatomy of that circuit itself.


Note that we have used linear Fisher information to quantify the population coding efficacy. Other information measures exist, and it is worth commenting on how much our findings generalize to different measures. In the case of jointly Gaussian stimulus and response distributions, correlations that maximize linear Fisher information also maximize Shannon's mutual information \cite{hu14}. In that regime our findings should generalize well. Moreover, whenever the neural population response distributions belong to the exponential family with linear sufficient statistics, the linear Fisher information is equivalent to the (nonlinear) ``full" Fisher information \cite{Beck:2011cb}. In practice, this is a good approximation to primary visual cortical responses to oriented visual stimuli~\cite{graf11,berens12},  and to other stimulus-evoked responses in other brain areas (see~\cite{moreno14} for discussion). Consequently, our use of linear Fisher information in place of other information measures is not a serious limitation.

For encoded sensory information to be useful, it must propagate from the periphery to the deep brain structures that guide behavior. Consequently, information should be encoded in a manner that is robust against corruption that arises during propagation. We showed that the features of population codes that maximize robustness can be substantially different from those that maximize the information content in peripheral layers. Moreover, by elucidating the set of covariances structures that optimize information transmission, we found that redundancy in a population code is neither necessary, nor sufficient, to guarantee robust propagation. In future work, it will be important to determine whether the nervous system uses the class of population codes that maximize information transmission.

Finally, while our main focus was on information propagation, the model we used -- linear feedforward weights followed by a nonlinearity -- is known to have powerful computational properties~\cite{hornik89}. It is, in fact, the basic unit in many deep neural networks. Thus, our main conclusion, which is that differential correlations are typically optimal, applies to any computation that can be performed by this architecture.

\section{Methods}

Here we provide detailed analysis of the relationship between correlations, feedforward weights, and information propagation. Our methods are organized as follows,
\begin{itemize}
\item Sec.~\ref{sec:iout}: we derive an expression for the information in the output layer (Eq.~\eqref{eq:LFI_y}). 
\item Sec.~\ref{sec:identify_optimal_family}: we identify the optimal family of first layer covariance structures when the gain function is linear.
\item Sec.~\ref{sec:opt_cov_nonlin}: we consider nonlinear gain functions.
\item Sec.~\ref{sec:info_loss}: we take a detailed look at the geometry of information loss.
\item Sec. ~\ref{sec:mininfo}: we prove that differential correlations minimize information.
\item Sec.~\ref{sec:variances_strategies}: we examine the relationship between coding strategies and information loss.
\item Sec.~\ref{sec:info_uu}: we compute information for a noise structure consisting of an arbitrary covariance matrix plus a rank 1 covariance matrix.
\item Sec.~\ref{numerical_details}: we provide details of our numerical simulations. 
\end{itemize}

\subsection{Information in the output layer}
\label{sec:iout}

Our analysis focuses on information loss through one layer of circuitry; to compute the loss, we need expressions for the linear Fisher information in the first and second layers.
Expressions for those two quantities are given in Eqs.~\eqref{eq:LFI_x} and \eqref{eq:LFI_y}.
The first is standard; here we derive the second.

To make the result as general as possible, we include noise inside the nonlinearity as well as outside it; if nothing else, that's probably a reasonable model for the spiking nonlinearity given in Eq.~\eqref{spike_noise}.
We thus generalize slightly Eq.~\eqref{eq:noisyY}, and write

\begin{align}
\label{y_zeta}
\yy = g(\ww \cdot \xx + \zzeta) + \eeta
\end{align}
where $\zzeta$ is zero mean noise with covariance matrix $\Sig_\zeta$, and here and in what follows we use the convention that $g$ is a pointwise nonlinearity, so for any vector ${\bf v}$, the $i^{\rm th}$ element of $g({\bf v})$ is $g(v_i)$. When $\Sig_\zeta = 0$, we recover exactly the model in Eq.~\eqref{eq:noisyY}.

Using Eq.~\eqref{eq:noisyX} for $\xx$, Eq.~\eqref{y_zeta} becomes

\begin{align}
\label{yi}
\yy = g \big( \hh (s) + \ww \cdot \xxi + \zzeta \big) + \eeta
\end{align}
where, recall, $\xxi$ and $\eeta$ are zero mean noise with covariance matrices $\Sigx$ and $\Sige$, respectively, and $\hh(s)$ is the mean drive to neuron $i$,

\begin{align}
\label{hdef}
\hh(s) \equiv \ww \cdot \ff(s).
\end{align}
To compute the linear Fisher information in the second layer, we start with the usual expression,

\begin{align}
\label{infoy}
I_y(s) &= \frac{\partial \text{E}({\yy}|s)}{\partial s}
\cdot \Cov[{\yy|s}]^{-1}
\cdot \frac{\partial \text{E}({\yy}|s)}{\partial s} 
\end{align}
where E and Cov denote mean and covariance, respectively.
The mean value of $\yy$ given $s$ is, via Eq.~\eqref{yi},

\begin{align}
\label{fydef}
\text{E}[\yy|s]
= \text{E}_{\xxi, \zzeta} \Big[ g \big( \hh(s) + \ww \cdot \xxi + \zzeta \big) \Big] \equiv
\overline{g} \big(\hh(s) \big)
.
\end{align}
Like $g(\cdot)$, $\overline{g}(\cdot)$ is a pointwise nonlinearity.
To compute the covariance, we assume, as in the main text, that $\xxi$ and $\eeta$ are independent; in addition, we assume that both are independent of $\zzeta$. Thus, the covariance of $\yy$ is the sum of the covariances of the first and second terms in Eq.~\eqref{yi}.
The covariance of the second term is just $\Sige$.
The covariance of the first term is harder. To make progress, we start by implicitly defining the quantity $\delta \Sig_g(s)$ via

\begin{align}
\label{cov_g}
\Cov\big[ g \big( \hh(s) + \ww \cdot \xxi + \zzeta \big) \big]
\equiv
\delta \Sig_g(s)
+
\left[
\weff(s) \cdot \Sigx \cdot \weff^T(s) + \GG'(s) \cdot \Sig_\zeta(s) \cdot \GG'(s) \right]
\end{align}
where $\weff(s)$ is the actual feedforward weight multiplied by the average slope of $g$,

\begin{align}
\label{weffdef}
W_{\text{eff},ij}(s) \equiv \overline{g}'\big( h_i(s) \big) W_{ij} ,
\end{align}
and $\GG'(s)$ is the a diagonal matrix with entries corresponding to the average slope of $g$,

\begin{align}
\label{Gdef}
\overline{G}'_{ij}(s) \equiv \overline{g}'\big(h_i(s)\big) \delta_{ij}
.
\end{align}
As in the main text, $\delta_{ij}$ is the Kronecker delta and a prime denotes a derivative.
The above implicit definition of $\delta \Sig_g$ is motivated by the observation that when $g$ is linear, $\delta \Sig_g$ vanishes.
Below, in Sec.~\ref{sec:nonlinear_C}, we show that if $\xxi$ is Gaussian, $\delta \Sig_g$ is positive semi-definite. Here we assume that the noise is sufficiently close to Gaussian that $\delta \Sig_g$ remains positive semi-definite, and thus can be treated as the covariance matrix of an effective noise source.
This last assumption is needed in Sec.~\ref{sec:opt_cov_nonlin}, where we argue that information loss is small when $\delta \Sig_g$ is small (see text following Eq.~\eqref{iy_dg}).

Making the additional definition

\begin{align}
\label{Sigeff_def}
\Sigeff(s) \equiv
\GG'(s) \cdot \Sig_\zeta(s) \cdot \GG'(s)
+ \delta \Sig_g(s)
+ \Sige
,
\end{align}
and using Eqs.~\eqref{yi} and \eqref{cov_g} and the fact that $\eeta$ is independent of both $\xxi$ and $\zzeta$, we see that

\begin{align}
\Cov[\yy|s] = \weff(s) \cdot \Sig_\xi \cdot \weff^T(s) + \Sigeff(s) .
\end{align}
Combining this with the expression for the mean value of $\yy$, Eq.~\eqref{fydef}, the linear Fisher information, Eq.~\eqref{infoy} becomes

\begin{align}
\label{iy}
I_y = \fprime \cdot \weff^T \cdot
\big[ \weff \cdot \Sigx \cdot \weff^T + \Sigeff \big]^{-1}
\cdot \weff \cdot \fprime
\end{align}
where we used Eqs.~\eqref{hdef} and \eqref{weffdef} to replace $\partial_s \text{E}(\yy|s)$ with $\weff \cdot \fprime$ and, to reduce clutter, we have suppresed any dependence on $s$. To pull the effective weights inside the inverse, we use the Woodbury matrix identity to write

\begin{align}
& \weff^T \cdot
\big[ \weff \cdot \Sigx \cdot \weff^T + \Sigeff \big]^{-1}
\cdot \weff
\\
\nonumber
& \ \ \ \ = \weff^T \cdot \Sigeff^{-1} \cdot \weff -
\weff^T \cdot \Sigeff^{-1} \cdot \weff \cdot
\big[ \Sigx^{-1} + \weff^T \cdot \Sigeff^{-1} \cdot \weff \big]^{-1}
\cdot \weff^T \cdot \Sigeff^{-1} \cdot \weff .
\end{align}
Then, using the fact that
$[{\bf A} + {\bf B}]^{-1} = {\bf A}^{-1} \cdot [{\bf A}^{-1} + {\bf B}^{-1} ]^{-1} \cdot {\bf B}^{-1}$, and applying a very small amount of algebra, this becomes

\begin{align}
& \weff^T \cdot
\big[ \weff \cdot \Sigx \cdot \weff^T + \Sigeff \big]^{-1}
\cdot \weff
\\
\nonumber
& \ \ \ \ = \weff^T \cdot \Sigeff^{-1} \cdot \weff \cdot
\left[ \id -
\Sigx \cdot \big[ \Sigx + (\weff^T \cdot \Sigeff^{-1} \cdot \weff)^{-1} \big]^{-1}
\right]
\end{align}
where $\id$ is the identity matrix.
It is then straightforward to show that

\begin{align}
\weff^T \cdot
\big[ \weff \cdot \Sigx \cdot \weff^T + \Sigeff \big]^{-1}
\cdot \weff
=
\big[ \Sigx + (\weff^T \cdot \Sigeff^{-1} \cdot \weff)^{-1} \big]^{-1}
.
\end{align}
Inserting this into Eq.~\eqref{iy}, we see that the right hand side of that equation is equal to the expression given in Eq.~\eqref{eq:LFI_y} of the main text.

\subsubsection{$\delta \Sig_g$ is positive semi-definite for Gaussian noise}
\label{sec:nonlinear_C}

To show that $\delta \Sig_g$ (defined implicitly in Eq.~\eqref{cov_g}), is positive semi-definite for Gaussian noise, we'll show that it can be written as a covariance. To simplify the analysis, we make the definition

\begin{align}
\cchi \equiv \ww \cdot \xxi + \zzeta .
\end{align}
With this definition,

\begin{align}
\label{delta_chi}
\delta \Sig_g = \Cov_{\cchi} \big[ g \big( \hh + \cchi \big) \big] - \GG' \cdot \Sig_\chi \cdot \GG'
\end{align}
where here and in what follows we are suppressing the dependence on $s$, $\Sig_\chi$ is the covariance matrix of $\cchi$, and $\GG'$ is defined in Eq.~\eqref{Gdef}. Because we are assuming that both $\xxi$ and $\zzeta$ are Gaussian, $\cchi$ is also Gaussian.

We'll show now that $\delta \Sig_g$ is equal to the covariance of the function $g ( \hh + \cchi ) - \GG' \cdot \cchi$. We start by noting that

\begin{align}
\label{cov_delta}
\Cov_{\cchi}\big[ g ( \hh + \cchi ) - \GG' \cdot \cchi \big] 
& =
\Cov \big[ g ( \hh + \cchi ) ]
- 2 \Cov\big[g ( \hh + \cchi ), \GG' \cdot \cchi \big]
+ \GG' \cdot \Sig_\chi \cdot \GG'
.
\end{align}
We'll focus on the second term, which is given explicitly by

\begin{align}
\label{cov_g_xi}
\Cov\big[g ( \hh + \cchi ), \GG' \cdot \cchi \big] =
\GG' \cdot \int d \cchi \, P(\cchi) \, \cchi \, g( \hh + \cchi) .
\end{align}
When $P(\cchi)$ is Gaussian, 

\begin{align}
P(\cchi) \cchi = - \Sig_\chi \cdot \frac{\partial}{\partial \cchi} \, P(\cchi)
.
\end{align}
Inserting this into Eq.~\eqref{cov_g_xi} and integrating by parts, we arrive at

\begin{align}
\Cov\big[g ( \hh + \cchi ), \GG' \cdot \cchi \big] =
\GG' \cdot \Sig_\chi \cdot \int d \cchi \, P(\cchi) \,
\frac{\partial}{\partial \cchi}
\, g( \hh + \cchi)
.
\end{align}
Using the fact that $\partial_{\cchi} g(\hh + \cchi) = \partial_\hh g(\hh + \cchi)$, the above expression becomes

\begin{align}
\Cov\big[g ( \hh + \cchi ), \GG' \cdot \cchi \big] =
\GG' \cdot \Sig_\chi \cdot
\frac{\partial}{\partial \hh}
\int d \cchi \, P(\cchi) \,
\, g( \hh + \cchi)
=
\GG' \cdot \Sig_\chi \cdot \GG'
.
\end{align}
where the second equality follows from the definition of $\GG$ (Eq.~\eqref{Gdef}.
Inserting this into Eq.~\eqref{cov_delta}, we see that the right hand side of Eq.~\eqref{cov_delta} is exactly equal to the right hand side of Eq.~\eqref{delta_chi}.
Thus, $\delta \Sig_g$ can be written as a covariance, and so it must be positive semi-definite.

\subsection{Identifying the family of optimal covariance matrices}
\label{sec:identify_optimal_family}

Here we address the question: what noise covariance matrix optimizes information transmission?
In other words, what covariance matrix $\Sig_\xi$ maximizes the information given in Eq.~\eqref{eq:LFI_y}?
That is hard to answer when $g$ is nonlinear, because in that case $\Sigeff$ depends on $\Sig_\xi$ via $\delta \Sig_g$ (see Eqs.~\eqref{cov_g} and \eqref{Sigeff_def}). In this section, then, we consider linear gain functions; in the next we consider nonlinear ones.
To make our expressions more readable, we generally suppress the dependence on $s$.

Our goal is to maximize $\iout$ with $\iin$ fixed. Using the definition of $\Sigy$ given in Eq.~\eqref{sigydef}, for linear gain functions the information in the second layer (Eq.~\eqref{eq:LFI_y}) is written

\begin{align}
\label{iydef}
\iout = \fprime \cdot
\big[ \Sigx + \Sigy \big]^{-1} \cdot \fprime.
\end{align}
We use Lagrange multipliers,

\begin{align} 
\frac{\partial}{\partial \Sigx} \Big[
\fprime \cdot [\Sigx + \Sigy]^{-1} \cdot \fprime
- \lambda \big( \fprime \cdot \Sigx^{-1} \cdot \fprime - I_x \big)
\Big] = 0,
\end{align}
where $\lambda$ is a Lagrange multiplier that enforces the constraint $\fprime \cdot \Sigx^{-1} \cdot \fprime = I_x$. Taking the derivative and setting it to zero yields

\begin{align} 
\label{lagrange}
[\Sigx + \Sigy]^{-1} \cdot \fprime \fprime \cdot [\Sigx + \Sigy]^{-1}
= \lambda \Sigx^{-1} \cdot \fprime \fprime \cdot \Sigx^{-1}
\, .
\end{align}
In deriving this expression we used the fact that the gain functions are linear, which implies that $\Sigy$ does not depend on $\Sigx$.
Multiplying by $\Sigx + \Sigy$ on both the left and right, we arrive at

\begin{align}
\fprime \fprime = \lambda 
[ \id + \Sigy \cdot \Sigx^{-1} ] \cdot \fprime
\fprime \cdot [ \id + \Sigx^{-1} \cdot \Sigy ]
.
\label{eq:fpfp}
\end{align}
This is satisfied when
\begin{align} 
\label{sigsprop}
\Sigy \cdot \Sigx^{-1} \cdot \fprime \propto \fprime
\, .
\end{align}
There are two ways this can happen,
\begin{subequations}
\label{yxi}
\begin{align}
\Sigy \cdot \Sigx^{-1} & \propto \id
\\
\Sigy \cdot \Sigx^{-1} & \propto \fprime \textbf{a}
\end{align}
\end{subequations}
where $\bf a$ is an arbitrary vector.
Combining these linearly, taking into account that $\Sigx$ is a covariance matrix and thus symmetric, and enforcing equality in Eq.~\eqref{eq:fpfp}, we arrive at
\begin{align} 
\label{siginv}
\Sigx^{-1} =
\frac{I_x}{\alpha \ieta} \Sigy^{-1} +
\frac{(\alpha-1) I_x}{\alpha \ieta^2} \Sigy^{-1} \cdot \fprime \fprime \cdot \Sigy^{-1}
+ \pp \cdot \oo^{-1} \cdot \pp
\end{align}
where $\ieta$ is the information the output layer would have if there was no noise in the input layer,
\begin{align}
\ieta(s) = \fprime(s) \cdot \Sigy^{-1} \cdot \fprime(s) 
\end{align}
(this is the same expression as in Eq.~\eqref{ieta}, it's repeated here for convenience), $\oo$ is an arbitrary symmetric matrix, $\pp$ is a projection operator, chosen so that $\pp \cdot \fprime =0$,
\begin{align}
\pp \equiv \id - \frac{\fprime \fprime}{\fprime \cdot \fprime} \, ,
\end{align}
and $\alpha$ is arbitrary (but subject to the constraint that $\Sigx$ has no negative eigenvalues).
Note that $\pp$ is a linear combination of the right hand sides of Eqs.~(\ref{yxi}a) and (\ref{yxi}b)), with ${\bf a} = \fprime$ in the latter equation.
It is straightforward to verify that when $\Sigx$ is given by Eq.~\eqref{siginv}, Eq.~\eqref{eq:fpfp} is satisfied.

To find an explicit expression for $\Sigx$, not just its inverse, we apply the Woodbury matrix identity to Eq.~\eqref{siginv}; that gives us

\begin{align} 
\label{sigma_xi_simplified}
\Sigx = \Sig_\alpha - \Sig_\alpha \cdot \pp \cdot
\big( \oo + \pp \cdot \Sig_\alpha \cdot \pp \big)^{-1}
\cdot \pp \cdot \Sig_\alpha
\end{align}
where

\begin{align} 
\Sig_\alpha \equiv
\left[ \frac{I_x}{\alpha \ieta} \Sigy^{-1} +
\frac{(\alpha-1) I_x}{\alpha \ieta^2} \Sigy^{-1} \cdot \fprime \fprime \cdot \Sigy^{-1}
\right]^{-1}
=
\frac{\alpha \ieta}{I_x}
\left[ \Sigy - \frac{(\alpha-1) \fprime \fprime}{\alpha \ieta} \right]
\, .
\end{align}
Inserting this into Eq.~\eqref{sigma_xi_simplified}, we arrive at

\begin{align} 
\label{general_optimal_noise}
\Sigx =
\frac{\alpha \ieta \Sigy}{I_x} +
\frac{(1-\alpha) \fprime \fprime}{I_x}
- \left( \frac{\alpha \ieta}{I_x} \right)^2 \Sigy \cdot \pp \cdot
\left( \oo + \frac{\alpha \ieta}{I_x} \pp \cdot \Sigy \cdot \pp \right)^{-1} 
\cdot \pp \cdot \Sigy
\, .
\end{align}
This is the same as Eq.~\eqref{optimal_noise} in the main text, except in that equation we let $\oo$ go to $\infty$, so we ignore the projection-related term. Ignoring that term is reasonable, as it just puts noise in a direction perpendicular to $\fprime$, and so has no effect on the information.

By choosing different scalars $\alpha$ and matrices $\oo$, a family of
optimal $\Sigx$ is obtained. These all have the same input information, $\iin$,
and the same output information, $\iout$, after corruption by noise.
An especially interesting covariance matrix is found in the limit $\alpha = 0$, in which case

\begin{align}
\Sigx = \frac{\fprime \fprime}{I_x} \, .
\end{align}
These are so-called differential correlations~\cite{moreno14}.
Importantly, the choice $\alpha = 0$ is the only one for which the optimal correlational structure is independent of the correlations in the output layer, $\Sigy$. Note that pure differential correlations don't satisfy Eq.~\eqref{sigsprop}. As such, they represent a singular limit, in the sense that $\Sigx$ in Eq.~\eqref{general_optimal_noise} satisfies Eq.~\eqref{sigsprop} with alpha arbitrarily small, but not precisely zero.

The other covariance that we highlight in the text is  found for $\alpha = 1$ and $\oo \to \infty$, in which case $\Sigx = \frac{\alpha I_{\eta}}{I_x} \Sigy$. This is the matched covariance case.
 
\subsection{Nonlinear gain functions}
\label{sec:opt_cov_nonlin}

We now focus on differential correlations, and determine conditions under which they are optimal for information propagation when the gain function, $g(\cdot)$, is nonlinear.
In this regime, the effective noise in the second layer (the second term in brackets in Eq.~\eqref{eq:LFI_y}) depends on $\Sigx$. This greatly complicates the analysis, and to make headway we need to reformulate our mathematical description of differential correlations. This reformulation is based on the observation that differential correlations correspond to trial-to-trial variability in the value of the stimulus, $s$~\cite{moreno14}. 
Consequently, the encoding model in the input layer can be written as a multi-step process,

\begin{subequations}
\label{diff_corr}
\begin{align}
s & = s_0 + \delta_s
\\
\xx & = \ff(s) + \xxi(s)
\\
\yy & = g\big( \ww \cdot \xx(s) + \zzeta \big) + \eeta(s)
.
\end{align}
\end{subequations}
Here $s_0$ is the value of the stimulus that is actually presented.
However, the neurons in the input layer, $\xx$, encode $s$ -- a corrupted version of $s_0$. This is indicated by Eq.~(\ref{diff_corr}a), which tells us that $s$ deviates on a trial-to-trial basis from $s_0$, with deviations that are described by a zero-mean random variable, $\delta_s$.

To see that this model does indeed exhibit differential correlations, we Taylor expand Eq.~(\ref{diff_corr}b) around $s_0$, yielding a model of the form

\begin{align}
\xx \approx \ff(s_0) + \fprime(s_0)\delta_s + \xxi(s_0) ,
\end{align}
for which the covariance matrix is

\begin{align}
\label{covx}
\Cov[\xx] = \Var[\delta_s] \fprime(s_0) \fprime(s_0) + \Cov[\xxi|s_0]
.
\end{align}
The first term corresponds to differential correlations.

Equations~(\ref{diff_corr}b) and Eq.~(\ref{diff_corr}c) correspond exactly to our previous model (Eq.~\eqref{eq:covs_green_blue}a). Consequently, the information about $s$ in the first and second layers are still given by Eqs.~\eqref{eq:LFI_x} and \eqref{eq:LFI_y} of the main text. However, we can't use those equations for the information about $s_0$.
For that, we focus on the variance of its optimal estimator given $\xx$, which we denote $\hat{s}_{0}$. Because of the Markov structure of our model ($s_0 \leftrightarrow s \leftrightarrow x$), we can construct $\hat{s}_{0}$ by first considering the optimal estimator of $s_0$ given $s$, and then the optimal estimator of $s$ given $\xx$. The variance of $\hat{s}_{0}$ given $\xx$ is then simply the sum of the variances of these two (independent) noise sources.

The optimal estimator of $s_0$ given $s$ is simply $s$, with conditional variance $\Var[\hat{s}_0(s) | s_0] = \Var[\delta_s]$. The optimal estimator of $s$ given $\xx$ is $\hat{s}(\xx)$, with variance  $\Var[\hat{s}(\xx)|s]$. Consequently,

\begin{align}
\Var[\hat{s}_0|s_0] = \Var[\delta_s] + \int ds \, P(s|s_0) \Var[\hat{s}(\xx)|s]
.
\end{align}
As usual, we approximate the variance of $\hat{s}(\xx)$ given $s$ by the linear Fisher information, yielding an approximation for the total Fisher information about $s_0$ given $\xx$,

\begin{align}
\label{iin_diff}
\frac{1}{\itot_x(s_0)} = \Var[\delta_s] + \int ds \, \frac{P(s|s_0)}{\iin(s)} \, .
\end{align}
Similarly, the Fisher information about $s_0$ given $\yy$ is approximated by

\begin{align}
\label{iout_diff}
\frac{1}{\itot_y(s_0)} = \Var[\delta_s] + \int ds \, \frac{P(s|s_0)}{\iout(s)} \, .
\end{align}
Note that we are slightly abusing notation here: above, $\iin(s)$ and $\iout(s)$ referred to the total information about the stimulus; now they refer to the information about the stimulus that is encoded in the first layer, which is different from the actual stimulus, $s_0$.
However, it is a convenient abuse, as it allows us to take over our previous results without introducing much new notation.

Our first step is to parametrize the covariance matrix, $\xxi$, and $\Var[\delta_s]$, in a way that ensures that the information in the first layer $\itot_x(s_0)$ remains fixed while we vary $\xxi$ and $\Var[\delta_s]$.
A convenient choice is

\begin{subequations}
\label{corr_diff}
\begin{align}
\Var[\delta_s] & = \frac{1}{\itot_x}
\int ds \, P(s|s_0)
\frac{\epsilon I_0(s)}{1+\epsilon I_0(s)}
\\
\Sigx(s) & =
\frac{1}{\itot_x} \frac{I_0(s) \Sig_0(s)}{1 + \epsilon I_0(s)} \, ,
\end{align}
\end{subequations}
where

\begin{align}
\label{i0}
I_0(s) \equiv \fprime(s) \cdot \Sig_0^{-1}(s) \cdot \fprime(s) .
\end{align}
Inserting Eq.~\eqref{corr_diff} into Eq.~\eqref{iin_diff}, we see that $\itot_x(s_0) = \itot_x$, independent of $\Sig_0(s)$.

The information in the second layer about $s$, $\iout(s)$, is given by Eq.~\eqref{eq:LFI_y}, with $\Sigeff$ given in Eq.~\eqref{Sigeff_def}. It is convenient to make the definition

\begin{align}
\label{sigyeff_def}
\Sigyeff \equiv (\weff^T \cdot \Sigeff^{-1} \cdot \weff)^{-1}
.
\end{align}
This is the analog of Eq.~\eqref{sigydef}, but for nonlinear gain functions.
It is clear from Eqs.~\eqref{Sigeff_def} and \eqref{cov_g} that $\Sigeff$ depends on $\Sigx$; consequently, it depends on $\epsilon$.

To maximize information with respect to $\epsilon$, we take a two step approach.
We write

\begin{align}
\label{iy_eps}
\iout(s;\epsilon, \epsilon_0) \equiv
\fprime^T(s)
\big[ \Sigx(s,\epsilon) + \Sigyeff(s,\epsilon_0) \big]^{-1} \fprime(s).
\end{align}
Here $\Sigx(s,\epsilon)$ and $\Sigy(s,\epsilon_0)$ are the same as in Eqs.~(\ref{corr_diff}b) and \eqref{sigyeff_def}; we have just made the dependence on $\epsilon$ explicit.
The two steps are to maximize first with respect to $\epsilon$, then with respect to $\epsilon_0$.
If the two maxima occurr in the same place, then we have identified the covariance structure that optimizes information transmission.

In the first step we differentiate $\itot_y(s;\epsilon, \epsilon_0)$ with respect to $\epsilon$. To simplify the expressions, we make the definition

\begin{align}
\label{sigtot_def}
\Sigtot(s, \epsilon, \epsilon_0) \equiv
\Sigx(s,\epsilon) + \Sigyeff(s,\epsilon_0)
.
\end{align}
Combining Eqs.~\eqref{iout_diff}, \eqref{corr_diff} and \eqref{iy_eps}, we have

\begin{align}
\frac{\partial}{\partial \epsilon}
\frac{1}{\itot_y(s_0;\epsilon, \epsilon_0)}
=  \frac{1}{\itot_x}\int ds \, P(s|s_0) \frac{I_0}{ \left( 1 + \eps I_0 \right)^2} + \int ds \, \frac{P(s|s_0)}{I^2_y}
\, \fprime \cdot \Sigtot^{-1} \cdot
\frac{\partial \Sigx(s,\epsilon)}{\partial \epsilon}
\cdot \Sigtot^{-1} \cdot \fprime
\end{align}
where we used the fact that for any square matrix ${\bf A}(x)$, $(d/dx) {\bf A}^{-1} = - {\bf A}^{-1} \cdot d {\bf A}/dx \cdot {\bf A}^{-1}$, and we suppressed much of the $s$, $\epsilon$ and $\epsilon_0$ dependence for clarity.
Using Eq.~(\ref{corr_diff}b) for $\Sigx(s,\epsilon)$, the derivative with respect to $\epsilon$ in the second term is straightforward,

\begin{align}
\frac{\partial}{\partial \epsilon}
\frac{1}{\itot_y(s_0;\epsilon, \epsilon_0)}
& =  \frac{1}{\itot_x}\int ds \, P(s|s_0) \frac{I_0}{ \left( 1 + \eps I_0 \right)^2} -
\int ds \, \frac{P(s|s_0)}{I^2_y} \, \fprime \cdot \Sigtot^{-1} \cdot
\frac{ I_0^2 \Sig_0}{\itot_x \left(1 + \eps I_0 \right)^2}
\cdot \Sigtot^{-1} \cdot \fprime  
\\
\nonumber
&= \frac{1}{\itot_x}\int ds \, P(s|s_0)
\frac{I^2_0}{ I^2_y \left( 1 + \eps I_0 \right)^2}
\left[ \frac{I^2_y}{I_0} - \fprime \cdot \Sigtot^{-1} \cdot \Sig_0 \cdot \Sigtot^{-1} \cdot \fprime \right]
\end{align}
Then, applying the definition $I_y(s) = \fprime \cdot \Sigtot^{-1} \cdot \fprime$ (see Eqs.~\eqref{iy_eps} and \eqref{sigtot_def}), and making the new definition 
%
%
%
\begin{align}
{\bf V} \equiv \fprime \cdot \Sigtot^{-1} \cdot \Sig_0^{1/2}
,
\end{align}
we arrive at the expression

\begin{align}
\label{diydeps}
\frac{\partial}{\partial \epsilon}
\frac{1}{\itot_y(s_0;\epsilon, \epsilon_0)}
= 
\frac{1}{\itot_x} \int ds \, P(s|s_0) \,
\frac{I_0^2}{\iout^2 (1+\epsilon I_0)^2 }
\,
{\bf V} \cdot
\left[
\frac{\Sig_0^{-1/2} \fprime \fprime^T \Sig_0^{-1/2}}{I_0}
- \id \right]
\cdot {\bf V} .
\end{align}

The right hand side of Eq.~\eqref{diydeps} is negative or zero if the term in brackets is negative semi-definite; that is, if all its eigenvalues are non-positive. Since the term in square brackets is a rank one matrix minus the identity, all but one of its eigenvalues are equal to -1. The remaining eigenvalue is 0, with corresponding eigenvector $\Sig_0^{-1/2} \cdot \fprime$ (see Eq.~\eqref{i0}). Thus, $\partial (1/\itot_y(s_0; \epsilon, \epsilon_0)/ \partial \epsilon) \le 0$, and $\itot_y(s_0; \epsilon, \epsilon_0)$ must have a global maximum at $\epsilon = \infty$.
If $g$ is linear, $\Sigyeff$ doesn't depend on $\epsilon_0$, and $\epsilon=\infty$ corresponds to pure differential correlations.
We have, therefore, recovered the $\alpha=0$ limit of Eq.~\eqref{general_optimal_noise}.

When $\epsilon = \infty$, $\Sigx$ vanishes, and so the expression for the information in the second layer simplifies considerably.
Combining Eqs.~\eqref{iout_diff} and (\ref{corr_diff}a), we have, in the $\epsilon \rightarrow \infty$ limit,

\begin{align}
\frac{1}{\itot_y(s_0;\infty, \epsilon_0)} = \frac{1}{\itot_x} + \int ds \, \frac{P(s|s_0)}{\iout(s; \infty, \epsilon_0)}
\end{align}
where

\begin{align}
\label{iy_dg}
\iout(s; \infty, \epsilon_0) = \fprime(s) \cdot \weff^T(s; \epsilon_0) \cdot
\big[ \Sige
+ \GG'(s) \cdot \Sig_\zeta(s) \cdot \GG'(s)
+ \delta \Sig_g(s; \epsilon_0) \big]^{-1}
\cdot \weff(s; \epsilon_0) \cdot \fprime(s) .
\end{align}
The latter equation follows by combining the fact that $\Sigx(s,\infty)=0$ (Eq.~(\ref{corr_diff}b)) with the definitions of $\Sigyeff$ and $\Sigeff$ (Eqs.~\eqref{sigyeff_def} and \eqref{Sigeff_def}, respectively).

The total information in the output layer is maximized when $\iout(s_0;\infty, \epsilon_0)$ is maximized.
That quantity depends on $\epsilon_0$ via $\Sigx(s,\epsilon_0)$, the noise covariance in the input layer.
As can be seen from Eq.~(\ref{corr_diff}b), larger $\epsilon_0$ implies smaller $\Sigx(s,\epsilon_0)$.
That has two effects.
First, when $\Sigx(s,\epsilon_0)$ is small enough, the covariance matrix $\delta \Sig_g$ becomes small (see Eq.~\eqref{cov_g}, and note that $\delta \Sig_g$ is positive definite, as shown in Sec.~\ref{sec:nonlinear_C}).
This tends to increase $I_y(s)$.
However, the effective tuning curves, $\weff(s; \epsilon) \cdot \ff(s)$, also depend on $\Sigx(s,\epsilon_0)$ (see Eq.~\eqref{weffdef}).
It is possible that increasing $\Sigx(s,\epsilon_0)$ modifes the tuning curves such that $\iout(s)$ increases.
Consequently, it is impossible to make completely general statements.

Nevertheless, we can identify two regimes. First, if there is no added noise in the output layer ($\eeta = \zzeta = 0$), then $\iout(s; \infty, \epsilon)$ goes to $\infty$ as $\epsilon_0$ goes to $\infty$, thus maximizing the total information.
This holds, however, only if the tuning curves are sufficiently dense relative to the steepness of the tuning curves; otherwise, the Fisher information is no longer a good approximation to the true information.
For smooth tuning curves this is generally satisfied, but it is not satisfied for the noise-free spike generating mechanism we consider in the main text (Eq.~\eqref{spike}), since for that nonlinearity $\fprime(s) = 0$ with probability 1. We expect, though, that in the absence of noise, this particular nonlinearity introduces an error that is $\mathcal{O}(1/n)$, implying that $\iout(s; \infty, \epsilon) \propto n^2$.
Numerical simulations (not shown) corroborated this scaling.
Thus, for sufficiently large populations, differential correlations are optimal for the noise-free spike-generating nonlinearity.
Note, though, that the thresholds must be chosen so that there are always both active and silent neurons; otherwise, in the limit that $\Sigx$ vanishes, the activity will contain no information at all about the stimulus.

The second regime is one in which the tuning curves have been optimized.
In this case, modifying the tuning curves by adding noise decreases information, and again differential correlations optimize information transmission.

To summarize, we have analyzed the scenario considered in Sec.~\ref{sec:DG_spikes} -- namely, the neural activities at the second layer, $\yy$, are given by a nonlinear function of the neural activities at the first layer, $\xx$, with noise added both before and after the nonlinearity. In this case, whether or not differential correlations in the first layer optimize information transmission depends on the details. 
They do if $g$ is linear, the tuning curves are optimal, or there is no added noise in the second layer and the tuning curves are sufficiently dense relative to the steepness of the tuning curves. If none of these are satisfied, however, differential correlations may be sub-optimal.

\subsection{Analysis behind the geometry of information loss}
\label{sec:info_loss}

Our goal in this section is to make more rigorous the geometrical arguments in Fig.~\ref{fg:skew_cartoons}. We start with the observation that, for Gaussian distributed neural responses, the 1 standard-deviation probability contours for the responses in the first layer (magenta ellipses in Fig.~\ref{fg:skew_cartoons}) are defined by

\begin{align}
\label{1sd}
\Delta \rr \cdot \Sigx^{-1} \cdot \Delta \rr = 1,
\end{align}
where $\Delta \rr \equiv \ff(s) - \rr$ represents fluctuations around the mean response to stimulus $s$. In two dimensions, which we'll focus on here, Eq.~\eqref{1sd} becomes

\begin{align}
\label{1sd_2d}
\frac{\Delta r_1^2}{\sigma_1^2} + \frac{\Delta r_2^2}{\sigma_2^2} = 1
\end{align}
where $\sigma_1$ and $\sigma_2$ are the lengths of the principal axes of the covariance ellipse (so $\sigma_1^2$ and $\sigma_2^2$ are the eigenvalues of $\Sigx$) and $\Delta r_1$ and $\Delta r_2$ are distances spanned by the magenta ellipses along those axes.

As shown in Fig.~\ref{fg:skew_cartoons}, the intersection between the magenta ellipse (the one defined in Eq.~\eqref{1sd_2d}) and the signal curve tells us the uncertainty in the value of the stimulus.
To quantify this uncertainty, we simply set $\Delta \rr$ to $\fprime(s) \Delta s_x$ (the subscript $x$ indicates that this is the uncertainty in the input layer), insert that into Eq.~\eqref{1sd_2d}, and solve for $\Delta s_x$. Defining $\theta$ to be the angle between $\fprime(s)$ and the long principal axis (see Fig.~\ref{fg:skew_cartoons}, and note that $\theta=0$ in panel B), and letting $\sigma_1$ correspond to the length of the ellipse's major axis (so $\sigma_1 > \sigma_2$), we have

\begin{align}
\label{fisher_eig}
|\fprime(s)|^2
\left[
\frac{\cos^2 \theta}{\sigma_1^2} + \frac{\sin^2 \theta}{\sigma_2^2}
\right]
= \frac{1}{\Delta s_x^2} .
\end{align}

The left hand side is the linear Fisher information in the first layer \cite{Sompolinsky:2001hh}, a fact that is useful primarily because it validates our (relatively informal) derivation.
More importantly, we can now see how \textit{iid} noise affects information.
The addition of \textit{iid} noise simply increases the eigenvalues by $\sigma^2$, so the ratio of the information in the output layer to that in the input layer is

\begin{align}
\label{info_ratio_1}
\frac{I_y}{I_x} = 
\frac{\Delta s^2_x}{\Delta s^2_y} = 
\frac
{\frac{\cos^2 \theta}{\sigma_1^2 + \sigma^2} + \frac{\sin^2 \theta}{\sigma_2^2 + \sigma^2}}
{\frac{\cos^2 \theta}{\sigma_1^2} + \frac{\sin^2 \theta}{\sigma_2^2}}
.
\end{align}
We can identify two limits. First, if $\theta=0$ (as it is in Fig.~\ref{fg:skew_cartoons}B), this ratio reduces to

\begin{align} 
\left. \frac{I_y}{I_x} \right|_{\theta = 0} = 
\frac{\sigma_1^2}{\sigma_1^2 + \sigma^2}
\, .
\end{align}
Second, if $\tan \theta \gg \sigma_2/\sigma_1$ (which essentially means the green line in Fig.~\ref{fg:skew_cartoons} intersects the covariance ellipse on the side, as in panel A, rather than somewhere near the end, as in panel B), the ratio of the informations becomes

\begin{align}
\label{info_ratio_2}
\left. \frac{I_y}{I_x} \right|_{\tan \theta \gg \sigma_2 / \sigma_1} \approx 
\frac{\sigma_2^2}{\sigma_2^2 + \sigma^2}
\, .
\end{align}
Because $\sigma_1 > \sigma_2$, the information loss is larger in the second case than in the first.
And the longer and skinnier the covariance ellipse, the larger the difference in information loss.
Thus, this analysis quantifies the geometrical picture given in Fig.~\ref{fg:skew_cartoons}, in which there is larger information loss in panel A (where $\theta > 0$) than in panel B (where $\theta = 0$).

\subsection{Minimum information}
\label{sec:mininfo}

Here we ask: what correlational structure minimizes linear Fisher information? To answer that, we use the multi-dimensional analog of Eq.~\eqref{fisher_eig},
\begin{align}
\label{info_models_sum}
I_x(s) = |\fprime(s)|^2 \sum_k \frac{\cos^2 \theta_k}{\sigma_k^2}
\end{align}
where $\sigma_k^2$ is the $k^{\rm th}$ eigenvalue of the noise covariance matrix and $\theta_k$ is the angle between $\fprime(s)$ and the $k^{\rm th}$ eigenvector \cite{Sompolinsky:2001hh}.
We would like to minimize $I_x(s)$ with respect to the angles, $\theta_k$, and the eigenvalues, $\sigma_k^2$. Without constraints, this problem is trivial: information is minimized by having infinite variances for the neural activities. To make the problem better-formulated, we add a constraint that prevents the optimization procedure from simply identifying that trivial solution.

We'll come to the constraint shortly, but first we'll minimize information with respect to the angles, $\theta_k$. That minimum occurs when the eigenvector corresponding to the largest eigenvalue is parallel to $\fprime(s)$; ordering the eigenvalues so that $\sigma_0^2$ is the largest eigenvalue, we have $\cos \theta_0=1$ and $\cos \theta_{k>0} = 0$. Consequently, the information at the minimum is
\begin{align}
\label{info_min}
I_x(s) = \frac{|\fprime(s)|^2}{\sigma_0^2} .
\end{align}

The next step is to minimize $I_x(s)$ with respect to the eigenvalues, subject to a constraint on the covariance matrix. We consider constraints of the form
\begin{align}
\label{constraint}
C(\sigma_0^2, \sigma_1^2, ...) \le C_0
\end{align}
where, to avoid the trivial solution (of infinite neural variances), $C$ is an increasing function of each of it's arguments: for all $k$,
\begin{align}
\label{constraint_condition}
\frac{\partial C(\sigma_0^2, \sigma_1^2, ...)}{\partial \sigma_k^2} \ge 0 .
\end{align}
Examples of $C(\sigma_0^2, \sigma_1^2, ...)$ are the trace of the covariance matrix (the sum of the eigenvalues) and the Frobenius norm (the square root of the sum of the squares of the eigenvalues).

Because of Eq.~\eqref{constraint_condition}, the information, Eq.~\eqref{info_min}, is minimized and the constraint, Eq.~\eqref{constraint}, is satisfied when all the eigenvalues except $\sigma_0^2$ are zero.
At this global minimum, the covariance matrix, $\Sigx$, displays purely differential correlations,
\begin{align}
\Sigx = \sigma_0^2 {\bf v}_0 {\bf v}_0 \propto \fprime(s) \fprime(s)
\end{align}
where ${\bf v}_0$ is the eigenvector associated with the largest eigenvalue. The last term in this expression follows because the above minimization with respect to the angles forced ${\bf v}_0$ to be parallel to $\fprime(s)$.
Thus, for a broad, and reasonable, class of constraints on the covariance matrix, differential correlations minimize information.


\subsection{Variances of neural responses, and robustness to added noise, for different coding strategies}
\label{sec:variances_strategies}

Throughout most of our analysis we focused on optimality of information transmission.
However, also important is how much information is transmitted at the optimum.
That's the subject of this section.
For simplicity we consider a linear gain function, which we set, without loss of generality, to the identity.
That allows us to use the analysis in Sec.~\ref{sec:identify_optimal_family}, and in particular Eq.~\eqref{general_optimal_noise}, which links the noise in the input and output layers.

Our starting point is the derivation of an expression for the ratio of the information in the output layer to that in the input layer.
To do that, we dot both sides of Eq.~\eqref{lagrange} by $\fprime$ on the left and right sides and solve for $\lambda$; we then do the same, except we dot with $\fprime \cdot \Sigy^{-1} \cdot [ \Sigx + \Sigy ]$ on the left and its transpose on the right.
This yields, after a small amount of algebra,
\begin{align}
\label{info_ratio}
\frac{I_y}{I_x} = \frac{I_\eta}{I_\eta + I_x} = \frac{1}{1 + I_x/I_\eta}
\end{align}
where $I_x$, $I_y$ and $I_\eta$ are given by Eqs.~\eqref{eq:LFI_x}, \eqref{eq:LFI_y} and \eqref{ieta}, respectively.
For information to be transmitted efficiently, $I_x$, the information in the input layer, must be small compared to $I_\eta$, the information associated with the added noise in the output layer.
Below, we investigate the conditions under which $I_x \ll I_\eta$, and thus when information loss is small.

Our strategy is to express $I_x/I_\eta$ in terms of the single neuron variability, quantified as the average variance -- something that has an easy interpretation.
We consider two cases: the weights are set to the identity ($\ww = \id$), and the weights are more realistic (each neuron in the input layer connects to a large number of neurons in the output layer).
The first case, identity weights, is not very realistic; we include it because it is much simpler than the second.

While the analysis is straightforward, it is somewhat heavy on the algebra, so we summarize the results here. We consider two extremes in the family of optimal covariance structures: the ``matched'' case ($\alpha=1$ in Eq.~\eqref{general_optimal_noise}, and, for simplicity, $\Omega = \infty$) and differential correlations ($\alpha=0$).
For matched covariances, near complete information transfer ($I_x \ll I_\eta$) requires the effective variance of the noise in the second layer to be small.
For identity feedforward weights, the effective variance in the input and output layers is about the same, so information loss is large.
However, identity feedforward weights are never observed in the brain; instead, each neuron in the input layer connects to a large number of neurons in the output layer.
Using $N_x$ and $N_y$ to denote the number of neurons in the input and output layers, respectively, and $K$ the average number of connections per neuron, the effective noise is reduced by a factor or $KN_x^2/N_y^2$ (see Eq.~\ref{tr_wt_w} below). Thus, if the number of neurons in the output layer is larger than the number in the input layer by a factor much larger than $K^{1/2}$, near complete information transmission is possible.
For pure differential correlations, the story is much simpler: so long as the number of neurons in both layers is large, and the added noise doesn't have a strong component in the $\ff'(s)$ direction, near complete information transmission always occurs.

\subsubsection{Identity feedforward weights}
\label{sec:ww} 

We'll first consider identity feedforward weight, $\ww = \id$.
We'll start with the matched covariance case. Using Eq.~\eqref{general_optimal_noise}, we have

\begin{align}
\label{sigx_a1}
\Sigx = \frac{I_\eta}{I_x} \, \Sige .
\end{align}
Taking the trace of both sides of this expression gives

\begin{align}
\label{info_loss_a1}
\frac{I_x}{I_\eta} =
\frac{\langle \sigma_\eta^2 \rangle}{\langle \sigma_x^2 \rangle}
\end{align}
where $\left< \sigma_x^2 \right>$ is the average variance of the input layer noise and $\left< \sigma^2_{\eta} \right>$ is the average variance of the added noise. 
If the added noise is on the same order as the noise in the input layer, information loss is high. Because of synaptic failures and chaotic dynamics, we expect the added noise to be substantial, implying that matching covariances is not an especially good strategy for transmitting information, in the case where $\ww = \id$.

Next we consider differential correlations ($\alpha=0$ in Eq.~\eqref{general_optimal_noise}),

\begin{align}
\Sigx = \frac{I_\eta}{I_x} \,
\frac{\fprime \fprime}{\fprime \cdot \Sige^{-1} \cdot \fprime}
\end{align}
where we used Eq.~\eqref{ieta} for $I_\eta$, with $\Sigy$ replaced by $\Sige$.
Taking the trace of both sides gives us

\begin{align}
\label{info_loss_a0}
\frac{I_x}{I_\eta} =
\frac{1}{N_x}
\frac{\fprime \cdot \fprime}{\langle \sigma_x^2 \rangle
\, \fprime \cdot \Sige^{-1} \cdot \fprime}
\, .
\end{align}
If the added noise doesn't have much of a component in the $\fprime$ direction, then $\fprime \cdot \Sige^{-1} \cdot \fprime$ is $\order(N_x)$. In this case, in the large $N_x$ regime, $I_x \ll I_\eta$, and (according to Eq.~\eqref{info_ratio}) information loss is small. In other words, for large neural populations, differential correlations allow small information loss even when the amount of added noise is large.

An especially instructive case is \textit{iid} noise added at the second layer. Using $\sigma_\eta^2$ for its variance, Eq.~\eqref{info_loss_a0} simplifies to 

\begin{align}
\frac{I_x}{I_\eta} =
\frac{1}{N_x} \frac{\sigma_\eta^2}{\langle \sigma_x^2 \rangle}
\, .
\end{align}
Consequently, for differential correlations and reasonably large neural populations, information loss is relatively small unless the variance in the second layer is \textit{much} larger than the average variance in the first layer (by about a factor of $N_x$) -- something that is not observed in the brain.

Although pure differential correlations can minimize information loss, they are not biologically realistic, as they do not display Poisson-like variability. That's because for differential correlations, the variance of neuron $i$ scales as $f_i'(s)^2$ rather than $f_i(s)$.
Fortunately, this can be fixed with very little information loss by adding Poisson-like variability in the input layer.
Doing so reduces the information only slightly: for the covariance structure given in Eq.~(\ref{eq:covs_green_blue}a), the information is

\begin{align}
I_x = \frac{I_0}{1 + \epsilon I_0}
\end{align}
where

\begin{align}
I_0 = \fprime \cdot \Sig_0^{-1} \cdot \fprime
\end{align}
is the information associated with the covariance matrix $\Sig_0$
(see Sec.~\ref{sec:info_uu}). That information is large whenever $\Sig_0$ doesn't contain much of a component in the $\fprime$ direction and $N_x$ is large. If these hold, the information in the input layer is approximately equal to $1/\epsilon$ -- exactly what it is for pure differential correlations.
Moreover, so long as $\Sige$ also doesn't contain much of a component in the $\fprime$ direction, information in the output layer is also close to $1/\epsilon$, and very little information is lost.
Thus, nearly pure differential correlations are biologically realistic and can lead to very small information loss.

\subsubsection{Realistic feedforward weights}
\label{sec:ww_realistic}

For realistic feedforward weights, $\ww$, we need to use $\Sigy$ rather than $\Sige$ in Eq.~\eqref{sigx_a1}, with $\Sigy$ given by Eq.~\eqref{sigydef}. (Note that because the gain function is the identity, $\weff = \ww$.)
We'll start, as above, with the matched covariance case. Taking the trace of both sides of Eq.~\eqref{sigx_a1}, but with $\Sige$ replaced by $\Sigy$, we have

\begin{align}
\frac{I_x}{I_\eta} =
\frac{\tr[\Sigy]/N_x}{\langle \sigma_x^2 \rangle}
\end{align}
where tr denotes trace and, as above, $N_x$ is the number of neurons in the input layer.
Using the fact that for any positive semi-definite square $n \times n$ matrix ${\bf A}$ (i.e., for any covariance matrix $\bf A$),

\begin{align}
\frac{\tr[{\bf A}^{-1}]}{n} \ge \frac{n}{\tr[{\bf A}]} ,
\end{align}
we have

\begin{align}
\label{info_loss_w_a1}
\frac{I_x}{I_\eta} \ge
\frac{1}{\langle \sigma_x^2 \rangle \, \tr[\Sigy^{-1}]/N_x}
=
\frac{1}{\langle \sigma_x^2 \rangle \,
\tr[\ww^T \cdot \Sige^{-1} \cdot \ww]/N_x} \, ,
\end{align}
with the second equality following from Eq.~\eqref{sigydef}.

To get a handle on the size of the trace term in the numerator, we note that it can be written

\begin{align}
\label{trace-identity}
\tr[\ww^T \cdot \Sige^{-1} \cdot \ww] =
\tr[\ww^T \cdot \ww]
\left< 1/\sigma_\eta^2 \right>_{W}
\end{align}
where, defining ${\bf v}_k$ to be the $k^{\rm th}$ eigenvector of $\Sige$, normalized so that ${\bf v}_k \cdot {\bf v}_k = 1$, and $\sigma_k^2$ to be its corresponding eigenvalue,

\begin{align}
\left< 1/\sigma_\eta^2 \right>_{W} \equiv
\frac{1}{\tr[\ww^T \cdot \ww]}
\sum_k \frac{{\bf v}_k \cdot \ww \cdot \ww^T \cdot {\bf v}_k}{\sigma_k^2}
\, .
\end{align}
To see that this really is a weighted average, note that because the ${\bf v}_k$ form a complete, orthonormal basis,

\begin{align}
\sum_k {\bf v}_k \cdot \ww \cdot \ww^T \cdot {\bf v}_k = \tr [\ww \cdot \ww^T] .
\end{align}
Inserting Eq.~\eqref{trace-identity} into Eq.~\eqref{info_loss_w_a1} gives us

\begin{align}
\label{e84_ratio_trace}
\frac{I_x}{I_\eta} \ge
\frac{1}{\langle \sigma_\eta^2 \rangle \langle 1/\sigma_\eta^2 \rangle_W}
\,
\frac{1}{\tr[\ww^T \cdot \ww]/N_x}
\,
\frac{\langle \sigma_\eta^2 \rangle}{\langle \sigma_x^2 \rangle}
\, .
\end{align}

This is similar to Eq.~\eqref{info_loss_a1}, except for two prefactors.
The denominator of the first prefactor lies between
$\langle \sigma_\eta^2 \rangle/\sigma_{\eta, \max}$ and
$\langle \sigma_\eta^2 \rangle/\sigma_{\eta, \min}$.
We'll assume this is $\order(1)$ (for \textit{iid} noise it is exactly 1), although we note that it's possible to make it either relatively large or relatively small.
The second prefactor is more interesting, as it is the sum of a large number of terms,

\begin{align}
\label{eq:tr_ww}
\frac{\tr[\ww^T \cdot \ww]}{N_x} =
\frac{1}{N_x} \sum_{i=1}^{N_y} \sum_{j=1}^{N_x} W_{ij}^2 
\end{align}
where $N_y$ is the number of neurons in the output layer.
To determine the size of the weights, we use that fact that

\begin{align}
\langle y_i \rangle = \sum_{j=1}^{N_x} W_{ij} f_j ,
\end{align}
and note that $\langle y_i \rangle$ and $f_i$ should be about the same size, on average.
Assuming that each neuron in the input layer connects, on average, to $K$ neurons in the output layer, it follows that $W_{ij}$ is nonzero with probability $K/N_y$. Consequently,

\begin{align}
\langle y_i \rangle =
\sum_j W_{ij} f_j \sim \frac{N_x K}{N_y} W_{\text{typical}} f_{\text{typical}}
\end{align}
where $W_{\text{typical}}$ and $f_{\text{typical}}$ are the typical sizes of the nonzero weights and the $f_j$, respectively.
To ensure that $\langle y_i \rangle$ and $f_i$ are about the same size, we must have

\begin{align}
W_{\text{typical}} \sim \frac{N_y}{N_x K} \, .
\end{align}
Inserting this into Eq.~\eqref{eq:tr_ww}, and using the fact that $W_{ij}$ is nonzero with probability $K/N_y$, we have

\begin{align}
\label{tr_wt_w}
\frac{\tr[\ww^T \cdot \ww]}{N_x} \sim \frac{(N_y/N_x)^2}{K}
\end{align}
This can be large if $N_y \gg N_x K^{1/2}$. Using this relationship in Eq.~\eqref{e84_ratio_trace}, we see that information loss can be small in the case of matched covariances, if there is sufficiently large divergence from the input to output layers.

What about differential correlations, $\alpha=0$? To understand information loss in this case, $\Sige$ is replaced by $\Sigy$ in Eq.~\eqref{info_loss_a0}, giving us

\begin{align}
\frac{I_x}{I_\eta} =
\frac{1}{N_x}
\frac{\fprime \cdot \fprime}{\langle \sigma_x^2 \rangle
\, \fprime \cdot \ww^T \cdot \Sige^{-1} \cdot \ww \cdot \fprime}
\end{align}
where we used Eq.~\eqref{sigydef} for $\Sigy$.
Here the logic is the same as it was in the previous section: so long as $\Sigy$ doesn't have a strong component in the $\fprime$ direction,
$\fprime \cdot \ww^T \cdot \Sige^{-1} \cdot \ww \cdot \fprime$ is $\order(N_y)$,
and, since $\fprime \cdot \fprime \sim \order(N_x)$, information loss is $\order(1/N_y)$.
Thus, with realistic feedforward weights, as with the identity case, differential correlations lead to very small information loss in large populations.

\subsection{Information in a population with a rank 1 perturbation to the covariance matrix}
\label{sec:info_uu}

In the analysis of nonlinear gain functions in Sec.~\ref{sec:DG_spikes}, it was necessary to construct a covariance matrix such that the information in the first layer was independent of $\epsilon_u$ and $\uu$.
For that we included a prefactor $\gamma_u$ in the definition of the covariance matrix, $\Sigx$ (see Eq.~\eqref{sigma_uu}). Here we determine how $\gamma_u$ should depend on $\epsilon_u$ and $\uu$.
Our starting point is an expression for the inverse of $\Sigx$.
As is straightforward to show, via direct substitution, that's given by

\begin{align}
\Sigx^{-1} =
\big( \gamma_u \big[ \Sig_0 + \epsilon_u \uu \uu \big] \big)^{-1} 
=
\frac{1}{\gamma_u} \left[
\Sig_0^{-1} -
\frac{\epsilon_u \Sig_0^{-1} \cdot \uu \uu \cdot \Sig_0}
{1 + \epsilon_u \uu \cdot \Sig_0^{-1} \cdot \uu}
\right]
.
\end{align}
Thus, the information in the input layer, $\fprime \cdot \Sigx^{-1} \cdot \fprime$, is given by

\begin{align}
\label{info_diff}
\fprime \cdot \Sigx^{-1} \cdot \fprime =
\frac{1}{\gamma_u} \left[
\fprime \cdot \Sig_0^{-1} \cdot \fprime -
\frac{\epsilon_u ( \fprime \cdot \Sig_0^{-1} \cdot \uu )^2}
{1 + \epsilon_u \uu \cdot \Sig_0^{-1} \cdot \uu}
\right]
.
\end{align}
To ensure that this information is independent of $\gamma_u$, we let

\begin{align}
\gamma_u =
\frac{1}{I_x}
\left[
\fprime \cdot \Sig_0^{-1} \cdot \fprime -
\frac{\epsilon_u ( \fprime \cdot \Sig_0^{-1} \cdot \uu )^2}
{1 + \epsilon_u \uu \cdot \Sig_0^{-1} \cdot \uu}
\right]
.
\label{eq:gamma_u}
\end{align}
Note that $\gamma_u$ depends on $s$ as well as $\epsilon_u$ and $\uu$.

\subsection{Details for Numerical Examples}
\label{numerical_details}

In this section we provide details for the numerical simulations for each relevant figure.

\begin{figure}[t!]
\begin{center}
\includegraphics[width=6in]{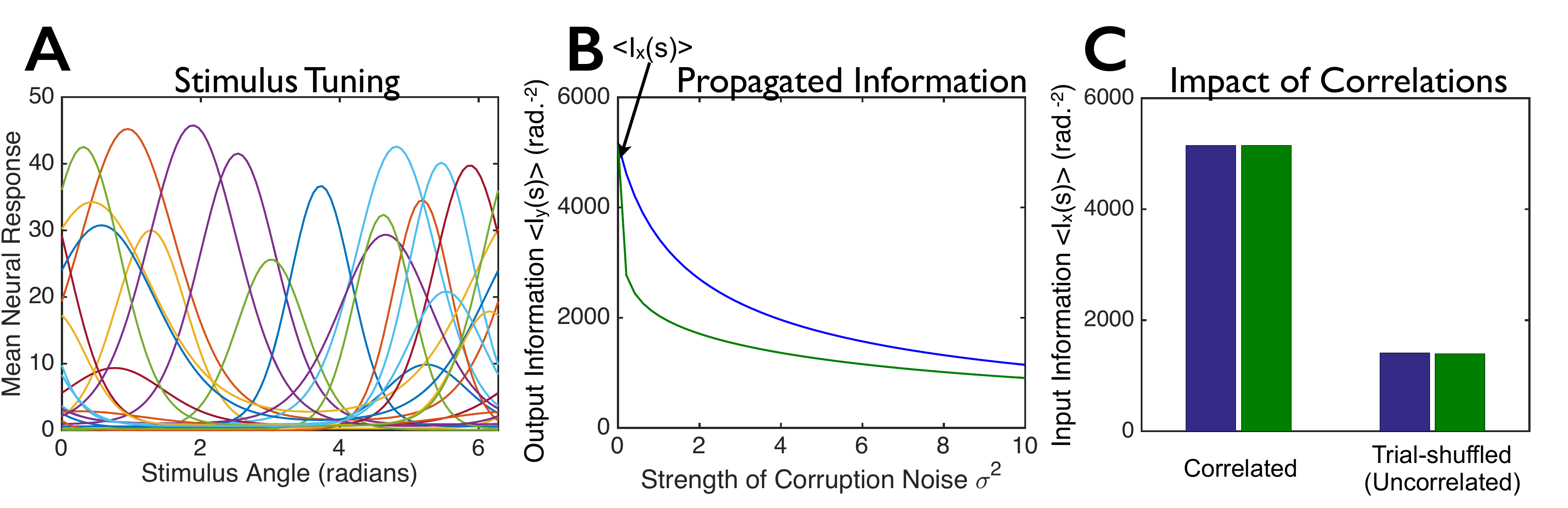}
\caption{ \textbf{Not all synergistic population codes are equally robust against corruption by noise.} This figure is similar to Fig.~\ref{fig:redundant}, but with synergistic instead of redundant population codes. We constructed two model populations -- each with the same 100 tuning curves (20 randomly-chosen example tuning curves are shown in panel \textbf{A}) -- for the first layer of cells. The two populations have different covariance structures $\Sig_{\xi}$ for their trial-to-trial variability (see main text, Eq.~\eqref{eq:covs_green_blue}), but convey identical amounts of information, $I_x(s)$, about the stimulus. (\textbf{B}) We corrupted the responses of each neural population by Gaussian noise (independently and identically distributed for all cells) of variance $\sigma^2$, to mimic corruption that might arise as the signals propagate through a multi-layered neural circuit, and computed the output information, $I_y(s)$, that these further-corrupted responses convey about the stimulus (blue and green curves). (\textbf{C}) Input information $I_x(s)$ in the two model populations (left; ``correlated") and information that would be conveyed by the model populations if they had their same tuning curves and levels of trial-to-trial variability, but no correlations between cells (right; ``trial-shuffled"). For panels B and C, we computed the information for 100 different stimulus values, equally spaced between 0 and $2 \pi$, and averaged the information over these stimuli. }
\label{fig:synergy}
\end{center}
\end{figure}

\subsubsection{Figure~\ref{fig:redundant} and its synergistic counterpart, Fig.~\ref{fig:synergy}}
\label{sec:fig2}

For the numerical examples in Fig. 2, we generated tuning curves for the first layer of cells using Von Mises distributions~\cite{Ecker:2011bx},

\begin{align}
\label{von-mises}
f_i(s) = \rho_i + \upsilon_i \exp\left[\beta_i \left(\cos(s- \phi_i) -1 \right) \right].
\end{align}
For each cell, the amplitudes, $\upsilon$, widths, $\beta$, peak locations, $\phi$, and baseline offsets, $\rho$, were drawn independently from uniform distributions with the following ranges,

\begin{itemize}
\item $\upsilon$:  1--51
\item $\beta$: 1--6
\item $\phi$: 0--2$\pi$
\item $\rho$: 0--1
\end{itemize}

\noindent
The covariance of the noise in the first layer was given by Eq.~\eqref{eq:covs_green_blue}, with the following parameters,

\begin{itemize}

\item blue population: $\epsilon = 10^{-3}$. 

\item green population: $\epsilon_u$ varies with stimulus so that, for each stimulus, the blue and green populations have identical information (on average, $\eps_u = 8 \times 10^{-3}$); $|\uu(s)| = |\fprime(s)|$; angle between $\uu(s)$ and $\fprime(s)$ = 1/8 of a radian.

\end{itemize}

\noindent
With these parameters, the two populations (blue and green) conveyed the same amount of information about the stimulus.

To rule out the possibility that differences in information robustness were due to differences in average correlations within the populations, we forced the average correlations to be the same for the blue and green populations. To do that, we repeatedly took random draws of the parameters describing the tuning curves ($\rho, v, \beta$ and $\phi$) until the population averaged correlations matched between the two populations. This resulted in average correlations of $-7 \times 10^{-5}$, and we used this set of tuning curves for our subsequent information calculations.

We computed the information, $I_y(s)$, in the second-layer responses using Eq.~\eqref{eq:LFI_y}, with $g(x) = x$, $\ww = \id$, and $\Sig_{\eta} = \sigma^2 \id$.
For the trial-shuffled information (Fig. 2C), we used Eq.~\eqref{eq:LFI_x}, with all off-diagonal elements of the covariance matrices $\Sig_{\xi}$ set to zero.
For all of these information calculations, we computed the information, $I_x(s)$ or $I_y(s)$, for 100 different stimulus values $s$, uniformly spaced between $0$ and $2 \pi$, and then averaged over these 100 different values.

To assess whether synergistic population codes can similarly vary in their robustness to corruption by noise, we repeated our calculations from Fig.~\ref{fig:redundant}, but modified the covariance matrices to make the population synergistic (Fig.~\ref{fig:synergy}C: the correlated responses convey more stimulus information than would independent cells with the same variances). To do that we again used the covariance matrices given in Eq.~\eqref{eq:covs_green_blue}, but we made $\epsilon$ and $\epsilon_u$ negative: $\epsilon=-5 \times 10^{-4}$ and $\left<\epsilon_u\right> = -3 \times 10^{-4}$ (as in Fig.~\ref{fig:redundant}, $\eps_u$ depends on the stimulus, $s$: it was chosen so that for each value of $s$ the blue and green populations have identical stimulus information). We chose $\uu(s)$ so that it had the same magnitude as $\fprime(s)$ and made an angle of 1/4 of a radian with $\fprime(s)$.
We used the same functions and distributions for the tuning curves as in Fig.~\ref{fig:redundant}, but used a different seed for the random number generator. As in Fig.~\ref{fig:redundant}, the seed was chosen (via multiple draws of the tuning curve parameters) so that the two populations had the same average correlations (in this case $2\times 10^{-5}$).
Also as in Fig.~\ref{fig:redundant}, the populations were roughly Poisson-like, in the sense that the mean and variance of the activity of each neuron was approximately equal. (Both the ``green" and the ``blue" populations have average Fano factors -- averaged over neurons and stimuli -- of 0.99.)
We again found that equally-informative population codes could vary significantly in terms of their robustness to noise (Fig.~\ref{fig:synergy}B).

\subsubsection{Figure~\ref{fg:fig_DG}}
\label{sec:fig5}

To generate Fig.~\ref{fg:fig_DG}B, we analytically computed the means of the second layer responses, resulting in the expression

\begin{equation}
\label{eq:DG_mean_out}
\mu_i(s) = \Phi \left[  \frac{f_i(s) - \theta_i}{\sigma_i(s)}    \right],
\end{equation}

\noindent where $\theta_i$ is the $i^{\rm th}$ cell's firing threshold, $\sigma_i$ is the standard deviation of the input noise to the cell, and $\Phi(\cdot)$ is the Gaussian cumulative distribution function. For each cell, the input function $f_i(s)$ was given by a Von Mises distribution, Eq.~\eqref{von-mises} (with the same distribution of parameters -- $v$, $\beta$, $\phi$ and $\rho$ -- as in the preceding examples), and the spiking threshold, $\theta_i$, was set to 3/4 of the peak height of the input tuning curve: $\theta_i = 3\left( \rho_i + \upsilon_i \right)/4$.

It is not straightforward to compute the covariance matrix of correlated responses generated by the dichotomized Gaussian model, so we used Monte Carlo methods to estimate the covariance: we took $10^6$ draws from the distribution of $\xx$, and for each draw we computed the corresponding responses, $\yy$, using the thresholding operation (Eq.~\eqref{spike}).  
We then computed the covariance of these simulated responses, and used them to estimate the linear Fisher information in the second layer activities via the standard expression,

\begin{equation}
I_y(s) = \frac{\partial \mmu(s)}{\partial s}
\cdot \Cov(\yy|s)^{-1} \cdot
\frac{\partial \mmu(s)}{\partial s}.
\label{eq:iy_meth}
\end{equation}

\subsubsection{Figure~\ref{fg:fig_DG_Zeta}}
\label{sec:fig6}

Figure~\ref{fg:fig_DG_Zeta} was made in the same fashion as Fig.~\ref{fg:fig_DG}, with the exception that noise was added before the spike generation nonlinearity. The noise, $\zzeta$, was Gaussian and drawn \textit{iid} with variance $\sigma_\zeta^2$.

\section*{Acknowledgments}
We thank Robert Townley, Kresimir Josic, Fred Rieke, Braden Brinkman, Maxwell Turner, and Alison Weber for helpful comments on the project.  JZ's contribution to this work was partially supported by an Azrieli Global Scholar Award from the Canadian Institute For Advanced Research (CIFAR). PEL was supported by the Gatsby Charitable Foundation. ESB acknowledges the support of NSF Grant CRCNS-1208027 and a Simons Fellowship in Mathematics, and thanks the Allen Institute founders, Paul G.~Allen and Jody Allen, for their vision, encouragement and support. 

\bibliography{papers}

\end{document}